# Deep learning-enhanced dual-mode multiplexed optical sensor for point-of-care diagnostics of cardiovascular diseases

*Gyeo-Re Han[1], Merve Eryilmaz[1,2], Artem Goncharov[1], Yuzhu Li[1], Shun Ye[2], Aoi Tomoeda[3], Emily Ngo[4], Margherita Scussat[2], Xiao Wang[1], Zixiang Ji[5], Max Zhang[6], Jeffrey J. Hsu[2,7], Omai B. Garner[8], Dino Di Carlo[2,9], and Aydogan Ozcan[1,2,9,10]**

[1]Electrical & Computer Engineering Department, [2]Bioengineering Department, [3]Chemical and Biomolecular Engineering Department, [4]Department of Psychology, [5]Department of Computer Science, [6]Department of Chemistry and Biochemistry, [7]Department of Medicine, [8]Department of Pathology and Laboratory Medicine, [9]California NanoSystems Institute (CNSI), [10]Department of Surgery, University of California, Los Angeles, CA 90095 USA.

**Corresponding Authors:* dicarlo@ucla.edu, ozcan@ucla.edu

## Abstract:

Rapid and accessible cardiac biomarker testing is essential for the timely diagnosis and risk assessment of myocardial infarction (MI) and heart failure (HF), two interrelated conditions that frequently coexist and drive recurrent hospitalizations with high mortality. However, current laboratory and point-of-care testing systems are limited by long turnaround times, narrow dynamic ranges for the tested biomarkers, and single-analyte formats that fail to capture the complexity of cardiovascular disease. Here, we present a deep learning-enhanced dual-mode multiplexed vertical flow assay (xVFA) with a portable optical reader and a neural network-based quantification pipeline. This optical sensor integrates colorimetric and chemiluminescent detection within a single paper-based cartridge to complementarily cover a large dynamic range (spanning ~6 orders of magnitude) for both low- and high-abundance biomarkers, while maintaining quantitative accuracy. Using 50 µL of serum, the optical sensor simultaneously quantifies cardiac troponin I (cTnI), creatine kinase-MB (CK-MB), and N-terminal pro-B-type natriuretic peptide (NT-proBNP) within 23 min. The xVFA achieves sub-pg/mL sensitivity for cTnI and sub-ng/mL sensitivity for CK-MB and NT-proBNP, spanning the clinically relevant ranges for these biomarkers. Neural network models trained and blindly tested on 92 patient serum samples yielded a robust quantification performance (Pearson's $r > 0.96$ vs. reference assays). By combining high sensitivity, multiplexing, and automation in a compact and cost-effective optical sensor format, the dual-mode xVFA enables rapid and quantitative cardiovascular diagnostics at the point of care.

## Introduction

Cardiovascular diseases (CVDs) remain the leading cause of death worldwide, with myocardial infarction (MI) and heart failure (HF) representing two of the most prevalent and interconnected conditions[1-5]. Approximately 25% of patients hospitalized with MI develop HF during admission or within 90 days after discharge[6], and ischemic heart disease from a prior MI is present in >40% of HF patients[7,8]. These conditions are pathophysiologically connected, often forming a vicious cycle: MI causes myocardial damage that leads to HF, and ischemic HF patients stay at elevated risk of recurrent infarction (**Fig. 1a**). In particular, co-diagnosis of MI and HF is associated with significantly worse outcomes, including higher mortality and rehospitalization rates, compared to patients with only one condition[6,8-11]. Together, MI and HF account for over one-third of cardiovascular mortality and impose major healthcare and socioeconomic burdens worldwide[4]. These bidirectional relationships highlight the importance of integrated diagnostic strategies. As demonstrated in recent clinical studies and guidelines[2,7,12,13], simultaneous assessment of MI and HF can provide clinically meaningful insights, enabling more precise patient stratification and earlier intervention.

Blood-based cardiac biomarkers are central indicators for diagnosing and risk-stratifying MI and HF (**Fig. 1a**). Cardiac troponin I (cTnI) is the gold-standard biomarker for MI[12]. It increases within 3–6 h after infarction and remains elevated for up to 10 days. Currently, high-sensitivity cTnI (hs-cTnI) assays are mainly used to detect changes in serial cTnI levels[14], which range from a few pg/mL to tens of ng/mL. Creatine kinase-MB (CK-MB), although largely replaced by cTnI for initial MI diagnosis, remains helpful for detecting reinfarction due to its shorter half-life. New symptoms appearing 72 hours to two weeks after the first MI, accompanied by CK-MB rising above 5 ng/mL, can indicate recurrent injury[15]. Meanwhile, N-terminal pro-B-type natriuretic peptide (NT-proBNP) reflects ventricular wall stress and is part of an HF guideline-recommended test panel. Clinical thresholds are set at ≥125 pg/mL for chronic HF and ≥300 pg/mL for acute HF, often reaching tens of ng/mL in severe cases[16]. Compared to B-type natriuretic peptide (BNP), another standard HF biomarker, NT-proBNP has a longer half-life and greater molecular stability, making it the preferred HF biomarker[17]. Each of these biomarkers is independently informative; however, emerging clinical evidence shows that testing them *simultaneously* provides a more comprehensive and clinically actionable profile for accurate diagnosis, risk stratification, and therapeutic guidance across MI, HF, and reinfarction[18-23].

Despite the value of comprehensive cardiac biomarker evaluation, current testing workflows are constrained by centralized laboratory testing[24], which is fragmented and slow, and remains misaligned with the latest clinical needs (**Fig. 1b**). For example, benchtop analyzers can process hs-cTnI, CK-MB, and NT-proBNP in parallel, but require separate cartridges, reagents, and large blood volumes (e.g., at least 55 μL of sample per test), increasing system complexity, labor, and cost. In addition to requiring substantial infrastructure and trained personnel, turnaround times often extend to several hours, particularly conflicting with the time-sensitive nature of MI, where early intervention within the first golden hour is critical[25]. Accessibility is also another major challenge[26], especially for patients in resource-limited or rural areas, as well as elderly and chronically ill individuals who face mobility challenges. In particular, patients recently hospitalized for acute HF require frequent clinic visits (e.g., four visits within the first two months) for titration of guideline-directed medical therapy guided by NT-proBNP levels[27]. These high-risk groups often visit the laboratory every 1–2 months for periodic biomarker monitoring to adjust medications, such as loop diuretics, as biomarker assessment has been shown to provide an additive benefit in patients with HF[28]. While point-of-care testing (POCT) offers a promising alternative by bringing rapid, cost-effective assays to the bedside, current systems suffer from serious limitations: most assays lack true high-sensitivity performance for troponin, multiplexing capabilities are limited, and devices are often still benchtop-sized rather than handheld (at-patient)[12,29]. In summary, existing centralized and POCT solutions fall short of delivering rapid, multiplexed, and sensitive cardiac biomarker assessment at the point of care (**Table 1**).

Here, we present a POCT-compatible optical sensor platform that addresses the growing demand for rapid, affordable, multiplexed, and high-sensitivity biomarker analysis (**Fig. 1c**). This optical sensor integrates a deep learning-enabled, dual-mode multiplexed vertical flow assay (xVFA) with a portable Raspberry Pi-based quantitative reader, combining chemiluminescence (CL) and colorimetric optical detection within a single paper-based cartridge. Using only 50 µL of serum, the platform *simultaneously* quantifies the concentrations of CK-MB, NT-proBNP, and cTnI within 23 min, enabling accurate, automated neural network-based quantification covering clinically relevant concentration ranges for each biomarker (**Fig. 1d**). The dual-mode optical sensing strategy employs CL for detecting cTnI with pg/mL-level sensitivity and colorimetry for robust detection of CK-MB, NT-proBNP, also achieving a large dynamic range, from sub-ng/mL to tens of ng/mL, thereby spanning the full diagnostic range for all three biomarkers. Our testing results demonstrate that the dual-mode xVFA achieves accurate biomarker quantification through neural network models trained, validated, and blindly tested using 92 patient serum samples. This dual-mode xVFA addresses the long-standing limitations of POCT by combining sensitivity, multiplexing capability, and automation in a single, compact format, enabling potential deployment across diverse healthcare settings, such as hospitals, emergency units, community clinics, and nursing homes. While the present work focuses on quantitative and multiplexing performance, the ability of our optical sensor to capture biomarker dynamics and cross-validate multiplexed signals establishes a foundation for integration of diagnosis, treatment monitoring, risk stratification, and prognosis into a unified, high-performance POCT platform for comprehensive testing of CVD and other diseases at the point of need.

## Results

### Design and workflow of the dual-mode optical sensor system

The dual-mode xVFA cartridge was engineered to combine sequential immunoassay and imaging steps within a compact, modular design (**Fig. 2a**). The 1st top cover allows for sample and reagent delivery and washing, while the 2nd top cover contains a reagent chamber and a transparent acrylic window for imaging. At the center of the cartridge, a multiplexed sensing membrane (12 mm × 12 mm) with 16 reaction spots enables parallel testing of CK-MB (3 spots), NT-proBNP (3 spots), and cTnI (4 spots), along with internal positive controls (2 spots) and negative controls (4 spots). Each testing spot was coated with specific anti-biomarker capture antibodies. Positive control spots were coated with secondary antibodies that broadly recognize and bind to anti-biomarker detection antibodies, providing a universal positive signal. In contrast, negative control spots were treated with buffer only.

The optimized paper layers in the 1st top case facilitate both vertical and lateral fluid transport, ensuring even and reproducible signal development across all assay spots in the sensing membrane (**Fig. S1**). Design optimization on the 2nd top included testing various cover materials and thicknesses to minimize imaging interference. A 1 mm thick acrylic cover showed better discrimination of low-concentration biomarkers than a 0.5 mm cover, while maintaining comparable CL performance, leading to its selection as the final component (**Fig. S2**). Details of this optimization are provided in *Supplementary Note 1*.

To support dual-mode detection, we developed a portable reader based on a Raspberry Pi device (**Fig. 2b**) that enables seamless acquisition of sensing membrane images in both colorimetric and CL modalities via a programmed graphical user interface (GUI). The system has a handheld footprint and combines an LED module with a cost-effective, high-resolution camera that offers a large dynamic range of exposure times (300 µs – 239 s) to cover typical exposure requirements for both shorter exposure colorimetric and longer exposure CL detection modes (see *Raspberry Pi-based dual-mode reader and image processing* sub-section in the Methods section for details). Colorimetric signals were imaged under green LED illumination to generate visible contrast from the light absorption by gold nanoparticles (AuNPs), while integrated CL signals were captured in dark conditions with the LED switched off to collect light generated by a luminol-based chemical reaction.

To prevent potential cross-interference between modalities, we developed a custom GUI to acquire colorimetric and CL images sequentially, without temporal overlap (**Fig. S3**). The GUI also allowed for manual adjustment of exposure times, enabling fine-tuning of the reader operation to achieve optimal sensitivity and performance. A custom cartridge tray was designed to accurately position the activated dual-mode xVFA cartridge inside a dark enclosure, minimizing misalignment and blocking environmental light intrusion during imaging. The captured images were processed to extract colorimetric and CL signals for each spot (see *Raspberry Pi-based dual-mode reader and image processing* sub-section in the Methods section for details), which were subsequently analyzed using neural network models for accurate biomarker quantification.

The rationale for implementing a dual-mode assay is to accommodate the distinct clinical ranges and diagnostic cut-off values of the three cardiac biomarkers (**Fig. 3a**). Specifically, cTnI must be quantified from the low pg/mL range up to tens of ng/mL to enable accurate monitoring of MI; NT-proBNP requires detection starting at hundreds of pg/mL for HF assessment; and CK-MB is clinically relevant at higher concentrations, typically in the ng/mL range, for reinfarction monitoring. Combining colorimetric (covering sub-ng/mL to higher ranges) and CL (covering the pg/mL range) detection in a single assay provides a practical solution and enables combined multiplexed analyses across all relevant concentration ranges, thereby broadening the applicability of POCT to diverse clinical scenarios where comprehensive risk assessment and prognosis are essential.

Each biomarker was paired with a specific conjugate design optimized for the corresponding detection modality, ensuring accurate quantification within its clinically relevant concentration range. CK-MB and NT-proBNP were detected using 40 nm AuNP-detection antibody conjugates for colorimetric readout (**Fig. 3b**), while cTnI was measured by a dual conjugate strategy: 15 nm AuNP coupled with polymerized horseradish peroxidase with streptavidin (PolyHRP-StA) and a biotinylated cTnI detection antibody (**Fig. 3c**), supporting both colorimetric (AuNP's intrinsic color, which is helpful at higher cTnI concentrations) and luminol-based CL signal generation (which is necessary for sub-pg/mL sensitivity). In particular, the PolyHRP-StA-based conjugate combines high antibody loading with a compact size (e.g., ~61 nm in hydrodynamic diameter), making it suitable for paper-based assays, and enhanced enzymatic turnover, which maximizes the CL signal output[30,31]. Its small diameter facilitates smooth transport through the submicron-scale porous channels of the paper matrix, ensuring uniform distribution and efficient immunoreaction within the stacked paper materials. It also offers straightforward preparation using centrifugation and nanoparticle-assisted quality control via simple spectral measurements (see the *Preparation of Conjugates* sub-section in the Methods section for details).

The overall workflow of the dual-mode xVFA is shown in **Fig. 3d**. The entire process takes less than 23 min, including 10 min for the immunoassay, 8 min for washing, and <5 min for imaging. The immunoassay begins by loading a mixture of serum (50 µL) and conjugate (50 µL) after wetting the membranes with running buffer. After incubation, the cartridge top is replaced with a new top case for washing to remove unbound reagents, ensuring that only the immuno-sandwich complex (capture antibody-target biomarker-detection antibody) is present at each reaction spot. The sensing membrane is then transferred to the imaging cartridge, where the CL substrate is applied. The portable optical reader automatically and sequentially captures colorimetric and CL images after 4 min of incubation. See the *Assay Operation* sub-section in the Methods section for details.

## Validation of cross-reactivity and assay performance

To evaluate potential cross-reactivity in the multiplexed optical sensor setup, we prepared serum samples spiked with various combinations of CK-MB, NT-proBNP, and cTnI within clinically relevant ranges (**Fig. 3e**, top). Both colorimetric and CL outputs were analyzed separately to assess specificity (**Fig. 3e**, bottom). Under optimized conditions (i.e., antibody concentrations and spot arrangement), elevated signals were observed only at the designated test spots for each biomarker, while all non-target spots remained at baseline noise level across all tested combinations. This observation confirms that the capture and detection antibody pairs for each biomarker functioned properly and independently, with minimal interference from co-existing analytes. Such testing was essential to demonstrate that multiplexing three cardiac biomarkers within a single cartridge does not compromise analytical specificity.

To assess analytical performance, we tested the dual-mode xVFA with spiked serum samples for each biomarker. The limit of detection (LoD) was determined according to the following definition: LoD = LoB + 1.645 × Standard Deviation (SD) of the lowest biomarker concentration measured[32]. Here, the LoB (limit of blank) was obtained as the mean blank value + 1.645 × SD of the blank samples, and concentrations were derived from the corresponding calibration curves. Detailed calculation procedures are provided in *Supplementary Note 2*.

For CK-MB, colorimetric signal-based calibration curves revealed a distinct concentration-dependent increase covering more than 2 orders of magnitude (**Fig. 3f**), with an LoD of 409 pg/mL (below the clinical cut-off of 5000 pg/mL) and a mean coefficient of variation (CV) of 3.0% across replicates. For NT-proBNP, the calibration curve showed a clear dose-dependent increase in colorimetric signals over 3 orders of magnitude (**Fig. 3g**), with an LoD of 40 pg/mL, well below the 125 pg/mL clinical threshold commonly used to rule out chronic HF. This level of sensitivity also suits the assessment of acute HF, where higher clinical thresholds (300 pg/mL) are used. The assay maintained good reproducibility, with a mean CV of 1.5% across tested concentrations.

For cTnI, we achieved quantification down to the sub-pg/mL range, reaching well below the clinical cut-off of <10 pg/mL used in high-sensitivity cTnI assays (**Fig. 3h**). Our complementary dual-mode sensing strategy enabled cTnI detection across a wide dynamic range spanning ~6 orders of magnitude. CL signals allowed reliable quantification down to an LoD of 0.12 pg/mL with a mean CV of 4.5% across replicates, while colorimetric signals ensured reliable cTnI quantification at higher concentrations ($\geq 10^4$ pg/mL), where CL optical signals became saturated and were no longer suitable for accurate quantification. Consistently, statistical comparisons (*t*-test) of spiked serum samples at low concentrations showed statistically significant differences ($P < 0.02$) in signal responses for all three biomarkers, confirming the assay's ability to detect clinically relevant changes near and below clinical cut-off levels (**Fig. S4**).

## Validation of dual-mode xVFA using patient serum samples

We analyzed 92 patient serum samples collected at UCLA Health for cTnI testing under an IRB-approved protocol (see the "Collection of patient serum samples and validation of biomarker reference concentrations" subsection in the Methods section for details). Ground truth concentrations for cTnI, CK-MB, and NT-proBNP were established using reference assays: cTnI was quantified with an FDA-cleared Access 2 immunoassay analyzer, CK-MB with a Luminex-based immunoassay, and NT-proBNP by an enzyme-linked immunosorbent assay (ELISA). The distributions of biomarker concentrations across the patient cohort are shown in **Fig. 4a**. For cTnI, 72 out of 92 patients (78.3%) had levels above the 4 pg/mL quantification threshold set by the standard analyzer. For CK-MB, 34 out of 92 patients (37.0%) had concentrations exceeding the 50 pg/mL threshold of the Luminex assay. For NT-proBNP, 59 of 92 patients (64.1%) had values above the 10 pg/mL quantification threshold of the ELISA.

Using these ground truth values, we classified the patient cohort according to clinically relevant cut-offs for different CVD subcategories. Joint analysis of cTnI and NT-proBNP concentrations enabled stratification of patients into distinct categories reflecting overlapping risks of MI and HF (**Fig. 4b**). Notably, 43 of 92 patients (46.7%) exhibited elevated levels of both biomarkers, indicative of compounded disease burden and highlighting the importance of integrated testing. Similarly, cross-classification of cTnI and CK-MB concentrations identified 9 cases (9.8%) with elevated CK-MB among patients having cTnI above the clinical threshold, consistent with either possible acute MI or potential reinfarction events (**Fig. 4c**). Moreover, in patients with chronically elevated troponin levels, a normal CK-MB result can help rule out acute ischemic injury, providing additional diagnostic reassurance about acute MI process. Together, this cross-classification analysis demonstrates that multiplexed biomarker testing can provide additional diagnostic insights beyond those accessible from single-marker measurements alone.

Next, we measured the 92 patient serum samples on the dual-mode xVFA to evaluate the correlation between the assay signals and the ground truth concentrations of each biomarker (**Figs. 4d–f**). For all three biomarkers, dual-mode xVFA signal intensities showed strong correlation with reference biomarker concentrations and demonstrated clear separation across clinically important ranges, including both the quantifiable thresholds established from spiking tests and concentrations above the clinical cut-offs. Quantifiable ranges were defined as follows: CK-MB ≥ 409 pg/mL, NT-proBNP ≥ 40 pg/mL, cTnI-CL ≥ 4 pg/mL, and cTnI-Color ≥ 1000 pg/mL. Within these ranges, correlation analyses using 4-parameter logistic (4-PL) curve fitting of the dual-mode xVFA signals demonstrated a very good agreement with ground truth values, with $R^2 = 0.990$ for CK-MB, 0.971 for NT-proBNP, 0.997 for cTnI-CL, and 0.984 for cTnI-Color. These results confirm that the dual-mode xVFA reliably captures clinically meaningful differences in biomarker levels across a heterogeneous patient cohort. Moreover, the dual-modality design enabled robust quantification over a wide dynamic range, consistent with serum spiking results reported above. Detailed 4-PL fitting parameters for each biomarker are provided in *Supplementary Note 3*. Notably, the measurements from these patient samples provided the foundational dataset for training compact neural network models that jointly analyze multimodal and multiplexed signal patterns. Leveraging the nonlinear relationships between colorimetric and CL optical responses, as well as among the three biomarkers, these compact network models enabled robust quantification of biomarker concentrations, as detailed in the next section.

## Neural network-based multiplexed quantification of CK-MB, NT-proBNP, and cTnI

Multiplexed quantification of biomarkers in clinical samples is a challenging task across many point-of-care assay platforms due to multiple sources of noise, including variability in sample flow, matrix effects from different patient serum conditions, and potential signal interference from cross-reactive domains[33-36]. To address these issues in our dual-mode xVFA, we implemented a deep learning-based approach[31,37-42], utilizing compact neural networks to establish functional relationships between the dual-mode xVFA signals (**Fig. 5a–d**) and the biomarker concentrations in patient serum samples. For each biomarker, we independently optimized a compact neural network-based quantification pipeline, resulting in a total of eight fully-connected neural network models: two for CK-MB ($DNN_{Class}^{CK-MB}$and $DNN_{\geq 500}^{CK-MB}$; **Fig. 5b**), two for NT-proBNP ($DNN_{Class}^{NT-proBNP}$ and $DNN_{\geq 125}^{NT-proBNP}$; **Fig. 5c**), and four for cTnI ($DNN_{Class}^{cTnI}$, $DNN_{<40}^{cTnI}$, $DNN_{40-1000}^{cTnI}$, and $DNN_{>1000}^{cTnI}$; see **Fig. 5d**). Full details of the model architectures are provided in the *Neural network-based CK-MB, NT-proBNP, and cTnI quantification pipeline* sub-section in the Methods section. All four models for CK-MB and NT-proBNP quantification used colorimetric signals as input (**Figs. 5b–c**), while the four cTnI models incorporated both colorimetric and CL signals (**Fig. 5d**) to increase sensitivity and extend the dynamic range (i.e., towards several tens of ng/mL level in upper clinical cut-off), covering ~6 orders of magnitude. For each biomarker, the quantification pipeline comprised one classification (*Class*) model and at least one quantification model.

All patient samples were first processed by the classification model of each biomarker, which classified them into predefined concentration ranges based on analytically and clinically informed classification thresholds. For CK-MB and NT-proBNP, classification thresholds were set below or at their respective clinical cut-offs (5000 pg/mL for CK-MB and 125 pg/mL for NT-proBNP), but close to the lower quantification limits of the dual-mode xVFA. This approach maximized quantification accuracy in the clinically critical transition region between normal and elevated levels while maintaining sufficient accuracy for higher biomarker ranges. For cTnI, the dual-mode xVFA exhibited sensitivity exceeding the clinical decision threshold (~40 pg/mL). Therefore, 40 pg/mL was set as the lowest classification threshold, aligning with the clinically established cut-off for myocardial injury and representing the concentration detectable above the assay's LoD in over 50% of healthy individuals, consistent with the analytical definition of a high-sensitivity assay[43]. A secondary threshold at 1000 pg/mL was established to distinguish the upper concentration range, where colorimetric signals complement CL-based quantification to avoid signal saturation.

In the CK-MB quantification pipeline, all samples were first processed by $DNN_{Class}^{CK-MB}$, which classified the samples into two concentration ranges: <500 pg/mL and ≥500 pg/mL (**Fig. 5b**). Samples assigned to the <500 pg/mL range were reported as CK-MB negative, since the clinical cut-off level for CK-MB is ~5000 pg/mL, approximately 10 times higher than the classification threshold. Only samples classified into the ≥500 pg/mL range were further processed by $DNN_{\geq 500}^{CK-MB}$ to measure the CK-MB concentration. Similarly, in the NT-proBNP quantification pipeline (**Fig. 5c**), samples were first classified into <125 pg/mL and ≥125 pg/mL ranges by $DNN_{Class}^{NT-proBNP}$. Samples in the higher range (≥125 pg/mL) were then quantified by $DNN_{\geq 125}^{NT-proBNP}$, while those below 125 pg/mL were reported as NT-proBNP negative. The 125 pg/mL threshold corresponds to the established clinical criterion for chronic HF (125 pg/mL), yet remains below the acute HF threshold (300 pg/mL), thereby enabling the model to effectively cover both chronic and acute HF cases within a unified framework.

The cTnI quantification pipeline (**Fig. 5d**) employed a three-range classification model ($DNN_{Class}^{cTnI}$) that categorized samples into <40 pg/mL, 40–1000 pg/mL, and >1000 pg/mL. The use of three ranges was necessary to accommodate the broad dynamic span of cTnI concentrations observed in clinical samples (4–21,000 pg/mL, spanning ~5 orders of magnitude), compared to the narrower 2–3 orders of magnitude observed for CK-MB and NT-proBNP concentrations. After this classification, samples were processed by one of the three range-specific cTnI quantification models ($DNN_{<40}^{cTnI}$, $DNN_{40-1000}^{cTnI}$, and $DNN_{>1000}^{cTnI}$), each independently trained to measure the biomarker concentration within its assigned range. Unlike the other biomarkers, no samples were directly reported as cTnI negative, as all the classification thresholds in $DNN_{Class}^{cTnI}$ exceeded the cTnI clinical cut-off, which stays in the 10–40 pg/mL range based on the 99th percentile upper reference limit[41]. Consequently, the final interpretation of cTnI negativity is left to the clinician's judgement, considering both the neural network-predicted cTnI value and the patient's medical record.

As part of an automated quality assurance process, if the biomarker concentration predicted by a quantification model contradicted the concentration range assigned by the corresponding classification model (i.e., predicted concentration fell > 5% outside the classification range boundaries), the sample was labeled "undetermined" for that biomarker and excluded from the quantification pipeline. Therefore, biomarker quantification only occurred when predictions from the classification and quantification models were in agreement with each other. Within each biomarker quantification pipeline, the neural network models operated in synergy to improve the accuracy of multiplexed biomarker concentration inference.

The architectures of all eight fully connected neural network models were independently optimized using the training sets of clinical samples (see the *Neural network-based CK-MB, NT-proBNP, and cTnI quantification pipeline* sub-section in the Methods section for details). In addition to optimizing these models' architectures, we also optimized the models' inputs by selecting an optimal subset of immunoreaction conditions using a feature selection algorithm[37,39,41], adjusted independently for each biomarker and network model. Feature selection was performed using an iterative backward feature elimination procedure, in which conditions were removed iteratively one at a time, starting from the full set. The total set of conditions includes five for CK-MB and NT-proBNP (i.e., three testing conditions [CK-MB, NT-proBNP, and cTnI], one positive control, and one negative control captured in the colorimetric mode) and six for cTnI (i.e., one testing condition [cTnI], one positive control, and one negative control captured in both the colorimetric and CL modes). Optimal condition subsets for the classification models were determined based on the highest training accuracy, while the optimal subsets for the quantification models were identified by minimizing the loss function, specifically, the root mean squared logarithmic error (RMSLE) loss for NT-proBNP and the mean squared error (MSE) for CK-MB and cTnI between the predicted and the ground truth biomarker concentrations (see **Fig. S5** for the optimal conditions selected for these eight neural network models).

As a result of this feature selection procedure, the CK-MB and NT-proBNP models used varying numbers of inputs, ranging from one to five (**Figs. 5b–d** and **Fig. S5**). Importantly, some of the classification and quantification models within CK-MB and NT-proBNP quantification pipelines incorporated non-target spot signals in addition to the biomarker-specific signal, suggesting that non-target spots may carry important information relevant to the biomarker of interest. For example, $DNN_{Class}^{CK-MB}$ and $DNN_{\geq 125}^{NT-proBNP}$ both used all three testing conditions, including both non-target spot conditions, alongside the specific testing condition (**Figs. 5b–c**). Within the cTnI pipeline, $DNN_{Class}^{cTnI}$ and $DNN_{>1000}^{cTnI}$ leveraged signals from both colorimetric and CL modes to capture the wide dynamic range of cTnI concentrations in clinical samples and mitigate saturation of CL signals at high concentrations (>1000 pg/mL). In contrast, $DNN_{<40}^{cTnI}$ and $DNN_{40-1000}^{cTnI}$ relied solely on CL signals, enabling greater sensitivity and more accurate cTnI quantification in the lower concentration ranges (**Fig. 5d**). Further details about the feature selection results and models' performance on the training sets are provided in **Fig. S5** and in *Feature selection* and *Neural network-based CK-MB, NT-proBNP, and cTnI quantification pipeline* sub-sections in the Methods section.

After optimizing the neural networks (including architectures and input features) on the validation training sets, we evaluated the final models' performance using blind testing samples – never seen before (**Fig. 6a**). These patient serum samples were processed with the same measurement and analysis protocols as those used for training and validation sets, but were completely new to the trained neural networks, providing an unbiased assessment of the models' ability to generalize to new patients. As shown in **Figs. 6b–d**, the classification networks ($DNN_{Class}^{CK-MB}$, $DNN_{Class}^{NT-proBNP}$, and $DNN_{Class}^{cTnI}$) demonstrated strong generalization to unseen samples, achieving accuracies of 95.5% (for CK-MB), 92.9% (for NT-proBNP), and 100% (for cTnI). Although some false negatives were observed, such as four for CK-MB and two for NT-proBNP, their clinical significance is expected to be minimal. The misclassified CK-MB samples (1000, 1042 [duplicate], and 3000 pg/mL) were 26–80% below the established clinical cut-off (5000 pg/mL), indicating low risk of diagnostic misinterpretation. The two NT-proBNP samples (221.5 and 308.3 pg/mL) were near the boundary between the chronic HF rule-out threshold (125 pg/mL) and the acute HF diagnostic cut-off (300 pg/mL), where follow-up or serial testing is routinely performed in standard clinical workflows to confirm disease progression. From a clinical standpoint, such results would be interpreted as non-critical findings prompting observation or confirmatory testing, rather than misdiagnosis. Therefore, these errors are not expected to alter patient triage or management decisions.

In the subsequent quantification analysis, CK-MB concentrations predicted by $DNN_{\geq 500}^{CK-MB}$ for 28 clinical samples classified into the ≥500 pg/mL range were strongly correlated with ground truth measurements from a commercial Luminex-based immunoassay (Pearson's $r$ = 0.966; **Fig. 6e**). Similarly, NT-proBNP predictions from $DNN_{\geq 125}^{NT-proBNP}$ for 33 samples in the ≥125 pg/mL range showed a very good agreement with the ELISA-based ground truth results (Pearson's $r$ = 0.986; **Fig. 6f**). Among the 33 NT-proBNP-positive samples, three were classified as false positives, with ground truth NT-proBNP concentrations staying in the <125 pg/mL range. In these cases, the dual-mode xVFA predictions from $DNN_{\geq 125}^{NT-proBNP}$ were around 300 pg/mL, approximately twice the ground truth values. Interestingly, a trend was observed where the classification accuracy appeared to correlate with the degree of automation in the ground truth assay: cTnI (100%), quantified by a fully-automated FDA-cleared analyzer, and CK-MB (95.5%), quantified by a semi-automated Luminex platform, exhibited higher classification accuracies than NT-proBNP (92.9%), whose ground truth relied entirely on manual ELISA testing, which is prone to errors. These findings suggest that the reliability of the ground truth measurement method can influence the apparent classification performance of machine learning models, particularly in low-concentration ranges. Finally, for cTnI, combined predictions from the three quantification models ($DNN_{<40}^{cTnI}$, $DNN_{40-1000}^{cTnI}$, and $DNN_{>1000}^{cTnI}$) across 68 blind testing samples also showed strong agreement with the ground truth concentrations measured by an FDA-approved analyzer (Pearson's $r$ = 0.988; **Fig. 6g**).

In terms of quantification precision, CK-MB and cTnI showed high reproducibility, with CVs of 8.6% and 7.9%, respectively. In contrast, NT-proBNP quantification had relatively lower precision (CV = 26.1%), likely reflecting variability in the manually processed ELISA testing used for its ground truth measurements, as discussed above. Future studies employing FDA-cleared, automated NT-proBNP assays as a reference would enable a more accurate assessment of the classification and quantification performance of the neural network models for NT-proBNP in the dual-mode xVFA. Additionally, two blind samples were labeled as "undetermined" and excluded from cTnI quantification due to conflicting results between the classification ($DNN_{Class}^{cTnI}$) and quantification ($DNN_{<40}^{cTnI}$) network models.

## Discussion

Our dual-mode optical xVFA sets a new standard among commercial and research-grade cardiac biomarker assays. As shown in **Table 1**, while commercial benchtop analyzers, such as Alinity I[44], ACCESS 2[45], and PATHFAST[46], can measure multiple cardiac biomarkers, they require parallel single-plex cartridges with separate reagents, sample loading, and incubation for each marker. This fragmented workflow requires a larger sample volume, increases assay time, and raises the cost per test. Most commercial POCT systems, such as i-STAT[47] and Atellica VTLi[48], also maintain a single-biomarker-per-cartridge format, limiting multiplexing capability and clinical interpretability. Research prototypes have explored integrating 2- to 4-plex detection on electrode[49-53], plasmonic[32,54,55], microfluidic[56-60], or paper-based sensors[39,61-71]; however, none offer comprehensive single-sample-drop multiplexing, wide clinical dynamic range coverage, portable devices, and a turnaround time of under 25 min. Many still require benchtop readers or cannot span the full clinically-relevant range, especially for cTnI (0.01–100 ng/mL).

The novelty of our dual-mode system, integrating colorimetric and CL optical sensing modalities within a single cartridge and reader, is further highlighted by comparisons with prior dual-mode platforms (**Table S1**). Dual-modality approaches have been primarily investigated in cancer and metabolic biomarker testing, using combinations of colorimetric, electrochemical, or fluorescence signals to extend linear ranges or enable self-validation of readouts. However, these implementations were largely solution-based or benchtop-oriented, required separate readers for each modality, and targeted only a single analyte. Typical operations required >40 min with complex assay steps, limiting the practical use for POCT. In contrast, our dual-mode xVFA uniquely addresses these gaps by consolidating both colorimetric and CL modalities within a single, self-aligned assay cartridge. The compartmented sensing membrane and vertical flow architecture ensure uniform sample distribution across test zones, minimizing cross-talk and enabling reliable multiplexing within a single device[72]. This configuration allows simultaneous multi-biomarker quantification through automated imaging by a portable reader within a 23-min workflow. The unified design eliminates optical alignment issues, minimizes user handling, and enables seamless, neural network-based, accurate biomarker classification and quantification. Together, these attributes advance dual-mode biosensing from a laboratory concept to a clinically applicable, low-cost, and scalable POCT solution. To the best of our knowledge, our dual-mode xVFA represents the first demonstration of dual-modality paper-based testing applied to a critical multiplexed CVD panel in a low-cost and POCT-compatible format.

The cTnI assay particularly illustrates the strength of the dual-modality testing. CL signals extended the quantifiable range into the sub-pg/mL regime, while the colorimetric signals enabled robust performance above $10^3$ pg/mL, where CL signals saturated. This complementary design expanded the overall measurable range to nearly 6 orders of magnitude in cTnI concentration (**Figs. 3h** and **4f**). Importantly, for cTnI, the dual conjugate strategy using AuNP–PolyHRP-StA supported both modalities in a single run, providing two outputs per spot within one assay (**Fig. 3c**). Neural network-based analysis further confirmed that combined colorimetric and CL inputs improved quantification at high cTnI levels (**Figs. 5d** and **6g**), ensuring reliable predictions across the full clinical spectrum from early myocardial injury to severe infarction.

The compact neural network-based pipelines outlined in our Results section enhanced the robustness of multiplexed quantification while mitigating overfitting to the training data. Given our limited sample size, we deliberately chose not to use a separate validation set because partitioning a small validation set might introduce substantial overfitting to the few validation samples. Instead, we prioritized maximizing the size and representativeness of the training data, ensuring reliable coverage of all concentration ranges and a good balance with an independent blind testing set in terms of sample distribution. As detailed in the Methods section, we also adopted several strategies, including L2 regularization, dropout, and early stopping, to prevent overfitting of our models. Ultimately, evaluating models' performance on independent, blind testing samples never used during the training process provides practical validation of the models in a scenario where the sample size is limited.

Traditional statistical data fitting-based curves consistently yielded a lower correlation with the ground truth biomarker concentrations compared to neural network-based models across all biomarkers (**Fig. S6**). In our assay, the optimal models represented power-fitting curves: a single power-fitting model for CK-MB and NT-proBNP, and three separate models for cTnI, each tailored to its specific concentration range (<40 pg/mL, 40–1000 pg/mL, or >1000 pg/mL). For a fair comparison, all power-fitting equations were trained on the same clinical samples used for the optimal neural network models. Across all three biomarkers, these data fitting-based curves showed lower correlation with the ground truth biomarker values, likely due to their reliance on single-feature inputs, making them highly susceptible to noise caused by flow heterogeneity, sample matrix effects, and user handling errors that often occur due to the low-cost sensor designs and non-ideal testing environments found in POCT, unlike those in a clinical laboratory. In contrast, neural network-based measurement pipelines enhanced the robustness of multiplexed biomarker quantification by processing multiple input features, capturing hidden correlations, and learning nonlinear relationships that are challenging for conventional parametric curve-fitting methods, such as logistic regression.

Proper selection of the model input features is also critical for achieving high biomarker quantification accuracy in clinical samples. For instance, when all immunoreaction conditions were used as input features, the classification accuracy of NT-proBNP and cTnI models dropped to 87.1% (from 92.9%) and 97.1% (from 100%), respectively (**Fig. S7**). Moreover, the predicted biomarker concentrations showed reduced correlation with ground truth values, with Pearson's $r$ decreasing to 0.907 (from 0.966), 0.979 (from 0.986), and 0.983 (from 0.988) for CK-MB, NT-proBNP, and cTnI, respectively **(Fig. S8**). Furthermore, $DNN_{>1000}^{cTnI}$ model failed to converge during training when all the signal conditions were included as input, underscoring the importance of the feature selection step outlined in our Results section. We note that the training and optimization of these compact neural network models is a one-time development effort, after which the final platform provides end-users with a streamlined sample-to-answer workflow. Together, these findings highlight the growing importance of neural network-assisted data processing as a vital tool for next-generation POCT analytics.

Importantly, neural network analysis not only improved the accuracy of individual biomarker quantification but also revealed the potential of how multi-biomarker patterns may inform patient status beyond single-analyte interpretation (**Fig. 4** and **Fig. 6**). As illustrated in **Fig. 4b–c**, concurrent elevation of cTnI with NT-proBNP or CK-MB identifies distinct subgroups of patients associated with overlapping risks of HF or recurrent myocardial injury. These insights would be missed with single-marker testing. Our results suggest that multiplexed, AI-assisted assays could support future studies aiming to link combined biomarker signatures with patient outcomes and guide specific treatment strategies. Furthermore, by enabling rapid, quantitative results at the point of care, our system can empower clinicians to perform near-real-time risk stratification and initiate appropriate interventions without waiting for central laboratory turnaround. This integration of multiplexed testing and neural network-based analysis thus represents a pathway toward faster, more informed cardiovascular decision-making in decentralized settings.

From a practical standpoint, the dual-mode optical xVFA offers substantial gains in efficiency and cost-effectiveness. Conventional benchtop analyzers require separate cartridges, reagents, and long turnaround times for each cardiac biomarker assay. In contrast, our multiplexed optical sensor design integrates all three biomarkers into a single cartridge, reducing the total assay time to 23 min (including all 3 biomarkers) and lowering the per-test cost of goods to $6.44 (**Table S2a**). Performing these assays using individual VFA cartridges would cost a total of $12.73 ($4.25 for cTnI, $3.37 for CK-MB, and $5.11 for NT-proBNP), indicating that the multiplexed testing approach reduces sensor cost by nearly 50% while also minimizing user intervention and sample volume needed. Antibodies account for approximately half of the cartridge costs, while 3D-printed components represent ~21%. Thus, further cost reductions are achievable through bulk antibody procurement and the transition to injection-molded cartridge fabrication, which can significantly reduce the total cost per test and improve scalability and accessibility.

Beyond per-test savings, the system architecture favors decentralized deployment. The Raspberry Pi-based optical reader can be assembled for $260 per device (**Table S2b**). Based on prototype-level cost estimations, the Raspberry Pi-based reader and dual-mode xVFA cartridges can be produced at a fraction of the cost of a single benchtop analyzer, enabling a substantially larger number of POCT packages to be distributed across multiple frontline healthcare sites, such as emergency departments, urgent care clinics, community clinics, and rural health centers, substantially expanding access to cardiac diagnostics. This scalable and resource-efficient deployment model could extend access to high-sensitivity, comprehensive cardiac testing in diverse settings without relying on centralized laboratory analyzers, highlighting the transformative potential of dual-mode optical xVFA for clinical workflows.

Despite these advances, several limitations should be noted. Our clinical validation involved 92 serum samples from a single site, emphasizing the need for larger multi-center studies to assess generalizability. Current work was limited to serum testing; future research will expand to whole-blood samples to better reflect clinical workflows. Furthermore, adapting the platform for finger-prick or upper-arm capillary blood collection (e.g., Tasso[73] or RedDrop[74]) could make sample collection easier and reduce patient burden. Such an approach would not only enhance usability in decentralized clinics but also enable at-home/remote cardiac biomarker monitoring, where patients at risk of MI or HF could be followed longitudinally with minimal disruption to their daily lives.

In summary, this work presents the first dual-modality, paper-based multiplexed optical assay for three critical cardiac biomarkers, fully implemented in a point-of-care format. By combining multiplexing capability, high sensitivity, broad dynamic range, affordability, and neural network-based analysis within a compact paper-based optical sensor and a handheld multimodal reader, this system establishes a new standard for CVD-POCT. Furthermore, beyond cardiac applications, the dual-mode optical xVFA framework is adaptable to other biomarker panels, offering a versatile approach to democratize high-performance, multiplexed biomolecular diagnostics in resource-limited settings.

## Materials and methods

**Preparation of conjugates**: For multiplexed detection, three separate AuNP–antibody conjugates were prepared, each optimized for its specific target biomarker. The detection antibodies used in this study were: anti- cTnI (1 mg/mL; 19C7cc, HyTest), anti-CK-MB (1 mg/mL; 70438, BioCheck Inc.), and anti-NT-proBNP (1 mg/mL; 13G12cc, HyTest).

For CK-MB and NT-proBNP detection via colorimetric methods, 40 nm AuNPs (1 mL; BBI Solutions) incubated with 100 mM borate buffer (100 µL, pH 8.5; Thermo Scientific) were functionalized with the respective detection antibodies (10 µg) through 1 h of incubation at room temperature (RT). Conjugation relied on passive adsorption of antibodies onto the AuNP surface, followed by 1 h surface blocking with bovine serum albumin (BSA; 20 µL, 10% w/w; Thermo Scientific). The conjugates were purified via three rounds of centrifugation (7,700 g, 20 min, 4 °C) using 10 mM borate buffer (pH 8.5) and resuspended in a conjugate dilution buffer containing 5% w/w trehalose (Sigma), 0.5% w/w protein saver (Toyobo), 0.2% v/v Tween 20 (Sigma), and 1% v/v Triton X-100 (Sigma) in 10 mM phosphate-buffered saline (PBS, pH 7.2; Thermo Scientific).

For cTnI detection with both colorimetric and CL modalities, 15 nm AuNPs (1 mL; BBI Solutions) were sequentially functionalized with PolyHRP-StA (Sigma) and a biotinylated anti-cTnI detection antibody. PolyHRP-StA (11 µL, 1 mg/mL) was first adsorbed onto the AuNPs in borate buffer, followed by BSA blocking and three centrifugation cycles (25,000 g, 30 min, 4 °C) for purification. A biotinylated detection antibody (12 µg; e.g., 4 µL of a 3 mg/mL biotinylated antibody) prepared using Sulfo NHS–biotin chemistry (EZ-Link™ Sulfo NHS-LC-LC-Biotin, Thermo Scientific) was incubated with the AuNP–PolyHRP conjugates for antibody labeling via streptavidin–biotin interactions. Residual binding sites were blocked with biotin–BSA (5 µL, 5 mg/mL; Thermo Scientific). The conjugate was washed three times with 10 mM borate buffer (pH 8.5) by centrifugation (25,000 g, 30 min, 4 °C) to remove unbound components. The final conjugate pellet was resuspended in 100 µL of storage buffer containing 10 mM Fe(II)-EDTA, 4% (w/w) trehalose, 0.1% (w/w) BSA, and 1% (v/v) Triton X-100 in PBS (10 mM, pH 7.2).

UV-Vis spectral analysis using a microplate reader (Synergy H1; BioTek) confirmed successful conjugation by a redshift in the absorbance peak and was used to determine the titer of conjugate concentrations. The final conjugates were stored at 4 °C at an 8.5 OD concentration until use.

**Preparation of sensing membrane**: The sensing membrane was prepared through four main steps: wax printing, heat treatment to melt the wax, antibody immobilization, and blocking. First, the patterned sensing membrane layout, which includes 16 reaction spots and 1 vent spot, was printed onto a 0.45 µm nitrocellulose (NC) membrane (Sartorius) using a wax printer (Xerox). The design featured a black background (waxed) and white spots (non-waxed). After printing, the membrane was baked at 120 °C for 55 s in a convection oven (Across International), allowing the wax to diffuse into the NC pore structure and creating distinct hydrophobic (waxed) and hydrophilic (non-waxed) regions. Up to 30 sensing membranes were arranged in a 6 × 5 grid with a 1 mm spacing and processed together in a single batch.

Following the formation of the hydrophilic regions, capture antibodies were deposited onto the designated spots on the NC membrane, taking care to avoid contact between droplets. For the test spots, 0.8 µL of each capture antibody was dispensed as follows: anti-cTnI capture antibody (1 mg/mL; 560cc, HyTest), anti-CK-MB capture antibody (0.5 mg/mL; 70225, BioCheck Inc.), and anti-NT-proBNP capture antibody (1 mg/mL; 15C4cc, HyTest). For the positive controls, 0.8 µL of anti-mouse IgG (50 µg/mL; Southern Biotech) and 10 mM PBS (pH 7.2) were spotted on the positive and negative control spots, respectively. Then, the membranes were dried at 37 °C for 10 min using the same convection oven previously described. Next, the membranes were immersed in 1% (v/v) BSA solution for 30 min at RT. After blocking, the membranes were dried again at 37 °C for 15 min. Lastly, the dried NC membrane batch sheets were trimmed to a uniform size of a single sensing membrane and affixed to the sensing membrane tray using double-sided adhesive tape.

**VFA cartridges and paper materials**: VFA cartridges were fabricated using a combination of 3D printing and laser cutting. The sensing membrane tray was printed with an Ultimaker S3 (Ultimaker) using PETG filament at a layer thickness of 150 µm, while the top and bottom cases were produced with a Form 3 printer (Formlabs) using gray resin at a layer resolution of 100 µm. The assay cartridge components (the bottom case, membrane tray, and 1st top case) were designed to compress the stacked paper layers to approximately 25% of their original thickness, ensuring uniform fluid flow and improved assay and washing efficiency.

For the bottom case used in the assay, five absorbent pads were positioned in the central region of the cartridge to ensure consistent fluid wicking. Furthermore, a flat-bottom version of the bottom case was designed for the imaging setup, replacing the absorbent pad stacking components with a solid surface. This configuration minimized fluid movement within the chamber and facilitated uniform and reliable CL signal acquisition.

The 1st top case, designed for immunoassay and washing steps, was constructed by sequentially layering paper components (absorption layer, flow diffuser, primary spreading layer, interpad, secondary spreading layer, and supporting layer). These layers were patterned using wax printing (e.g., flow diffuser and supporting layer), cut to defined dimensions using a Trotec $CO_2$ laser cutter (e.g., absorption layer, primary spreading layer, interpad, and secondary spreading layer) or blade (e.g., flow diffuser and supporting layer), and bonded with adhesive foam tape. To reduce non-specific adsorption, selected paper layers (flow diffuser, interpad, and supporting layer) were pretreated with 1% (v/v) BSA solution.

The 2nd top case, designed for imaging the assay result, was fabricated from transparent acrylic (16.3 × 16.3 × 1 mm$^3$), laser-cut, and fixed onto the 3D-printed body with clear epoxy adhesive. The acrylic window contained a ventilation port (0.7 mm diameter) to facilitate air displacement during reagent loading. This chamber can hold up to 520 µL of CL substrate and was sealed via foam tape (underneath the sensing membrane), aligned with the extruded cartridge edges, effectively preventing leakage and evaporation without requiring additional gaskets.

**Raspberry Pi-based dual-mode optical reader and image processing**: The dual-mode optical reader had a portable form factor and used a Raspberry Pi computer paired with an inexpensive off-the-shelf camera module to capture both colorimetric and CL modalities of the dual-mode xVFA. To support the colorimetric modality, the optical reader incorporated green LEDs (DigiKey) with a central wavelength of 525 nm, matching the peak absorption range of AuNPs used in this work (i.e., 520–530 nm). The LEDs were arranged in a circular shape around the camera and polished from the front to provide a uniform illumination over the entire sensing membrane. The camera module consisted of a Raspberry Pi HQ camera (Adafruit) coupled to an M12-mount lens (Amazon), accommodating a broad range of exposure times from 300 µs to 239 s. The Raspberry Pi computer, camera module, LEDs, and a custom-designed LED driver were housed within a 3D-printed case produced by an Object 30 printer (Stratasys). The reader was also equipped with posts for pedestal installation, allowing both hand-held and benchtop use.

We prioritized shorter exposure times for the colorimetric modality to reduce background noise from the illumination LEDs[42]. In contrast, for the CL modality, we used longer exposure times to accumulate faint CL signals from low analyte concentrations over an extended time period, enhancing sensitivity[31]. Consequently, we chose exposure times of 800 µs and 30 s for the colorimetric and CL modalities, respectively. An intuitive GUI allowed users to select between the two modalities and featured input fields with user-adjustable parameters (i.e., exposure time, number of images to capture, and capture interval between subsequent images) for automated imaging. Additionally, the GUI enabled automated capturing of both colorimetric and CL images in a single run using the dual-mode regime. Importantly, during the dual-mode regime, the colorimetric and CL images were captured sequentially to prevent interference between the modalities. All images were acquired in raw format with consistent imaging settings.

After capturing colorimetric and CL images of the activated sensing membrane, the 16 immunoreaction spots were segmented from the image and pixels within each spot were averaged to generate 16 colorimetric and 16 CL intensity values (i.e., $s_{i,j}^{m}$, $i \in$ {CK-MB, NT-proBNP, cTnI, positive control [(+)Ctrl], and negative control [(-)Ctrl]} is the type of immunoreaction condition, $j$ is the spot repeat within the given condition, and $m \in$ {Color, CL} is the type of optical modality. Colorimetric intensities were calculated using a custom segmentation code that utilized the green channel of the original RGB image, and CL intensities were obtained using ImageJ software, which employed the blue channel. These signals were further normalized by the corresponding background signals ($b_{i,j}^{m}$) from the images of the sensing membrane captured before the dual-mode xVFA operation, and the final absorption signals for each image were calculated as:

$$X_{m}^{i,j} = 1 - \frac{s_{i,j}^{m}}{b_{i,j}^{m}}. \quad (1)$$

Alike spots within each immunoreaction condition were averaged (along $j$), generating a total of 5×2=10 absorption signals ($\bar{X}_{m}^{i}$), where the factor of 5 represents the five conditions (CK-MB, NT-proBNP, cTnI, positive control, and negative control) and the factor of 2 corresponds to the two optical modalities (colorimetric and CL). These signals were processed by compact neural network models (detailed below) to quantify biomarker concentrations in patient serum samples.

**Assay operation**: The dual-mode xVFA workflow consisted of two sequential phases: immunoassay and imaging. To start the assay, 200 µL of running buffer (0.5% v/v Triton X-100, 1% v/v Tween-20, and 1% v/v BSA in 10 mM PBS, pH 7.2) was applied to activate the top/bottom cases of the assembled xVFA and incubated for 30 s. The conjugate solution contained a mixture of three nanoparticle–antibody conjugates, diluted in conjugate buffer to achieve final concentrations of 2 OD for the CK-MB assay, 2 OD for the NT-proBNP assay, and 1.2 OD for the cTnI assay. A 100 µL reaction mixture, prepared by combining 50 µL of the conjugate mixture solution with 50 µL of the serum sample, was then added and incubated for 1.5 min. This step was followed by the addition of 350 µL of running buffer to maintain flow through the cartridge, facilitating immunoreactions and removing unbound conjugates. After an 8-min incubation, the assay top case was replaced with a washing top case, and 500 µL of running buffer was applied for another 8-min of washing.

For imaging, the sensing membrane tray was transferred to the imaging cartridge. A total of 440 µL of CL substrate solution (SuperSignal™ ELISA Pico Chemiluminescent Substrate, 37069, Thermo Scientific) was added, and the cartridge was inserted into the reader for automated dual-mode imaging. The reader first captured colorimetric images under green LED illumination. In the same configuration, CL imaging was then performed after a 4-min incubation with the LED turned off. All images acquired were processed using an image analysis pipeline to extract and quantify the signal intensities.

Through iterative optimization with over 400 test cartridges, the total assay time was standardized to 23 min. The immunoassay phase, which included incubation and washing, required 18 min to ensure complete binding and effective removal of unbound conjugates, followed by a 5-min step for CL reaction and image acquisition. Computational image analysis was completed within 1 s per cartridge, and biomarker concentrations were inferred by trained compact neural networks in less than 0.5 s, making the processing time negligible compared to the overall assay duration.

**Preparation of biomarker-spiked serum samples**: For assay optimization and validation, standard antigens for the three target biomarkers were used: human-derived cTnI (I-T-C complex; Lee Biosolutions), CK-MB protein (Biosynth), and recombinant human NT-proBNP (Hytest). Stock solutions were dispensed into 1 µL or 2 µL aliquots and stored under recommended conditions (cTnI and NT-proBNP at -80 °C; CK-MB at -20 °C) as specified by the manufacturers. Prior to use, aliquots were thawed once and used immediately to prevent degradation. Matrix selection for spiking experiments was based on baseline antigen levels measured by ELISA. cTnI-free serum (Hytest) and NT-proBNP-free plasma (Hytest) both contained negligible levels of cTnI and NT-proBNP (<0.5 pg/mL). CK-MB concentrations were 56.1 pg/mL in the cTnI-free serum and 251.1 pg/mL in the NT-proBNP-free plasma, both far below the clinical cut-off of 5000 pg/mL. Given the lower background levels of CK-MB, the cTnI-free serum was selected as the final matrix for spiking and serial dilutions in all validation studies.

**Collection of patient serum samples and validation of biomarker reference concentrations**: Clinical serum samples were collected under an IRB protocol from UCLA Health (IRB #20-002084). These samples are leftover/existing specimens collected separately from this study, and our results are not used for clinical decisions related to the anonymized patients. A total of 92 patient samples were analyzed. The ground truth concentrations of the three biomarkers were measured using established assays: high-sensitivity cTnI analyzer (Access 2, Beckman Coulter; quantification cut-off of 4 pg/mL), Luminex-based immunoassay for CK-MB (ProcartaPlex, Thermo Fisher; quantification cut-off of 50 pg/mL), and ELISA for NT-proBNP (Abcam; quantification cut-off of 10 pg/mL). Values below each assay-specific cut-off were reported as "<cut-off concentration" (e.g., <4 pg/mL for cTnI), while samples at or above the threshold were measured quantitatively.

Among the 92 patient samples, cTnI was <4 pg/mL in 20 cases and ≥4 pg/mL in 72 cases; CK-MB was <50 pg/mL in 58 cases and ≥50 pg/mL in 34 cases; NT-proBNP was <10 pg/mL in 33 cases and ≥10 pg/mL in 59 cases. All clinical samples were stored at -80 °C and, prior to testing, thawed at 4 °C and equilibrated to RT immediately before use.

**Neural network-based CK-MB, NT-proBNP, and cTnI quantification pipeline**: The neural network-based quantification pipeline for each of the three biomarkers (i.e., CK-MB, NT-proBNP, and cTnI) comprised a two-stage process: (1) classification of the biomarker concentration range; and (2) quantification of the biomarker concentrations within different ranges. We independently optimized compact neural network models for each quantification pipeline, converging to a total of 8 network models: $DNN_{Class}^{CK-MB}$ and $DNN_{\geq 500}^{CK-MB}$ for CK-MB; $DNN_{Class}^{NT-proBNP}$ and $DNN_{\geq 125}^{NT-proBNP}$ for NT-proBNP; and $DNN_{Class}^{cTnI}$, $DNN_{<40}^{cTnI}$, $DNN_{40-1000}^{cTnI}$ and $DNN_{>1000}^{cTnI}$ for cTnI. All models in the CK-MB and NT-proBNP quantification pipelines used the colorimetric signals from dual-mode xVFA as input, while all four models in the cTnI quantification pipeline utilized both the colorimetric and CL signals as input data.

In the first stage, as shown in **Fig. 5**, a set of three classification neural networks ($DNN_{Class}^{CK-MB}$, $DNN_{Class}^{NT-proBNP}$, and $DNN_{Class}^{cTnI}$) simultaneously classified each sample under test into predefined concentration ranges: <500 pg/mL or ≥500 pg/mL for CK-MB (using $DNN_{Class}^{CK-MB}$); <125 pg/mL or ≥125 pg/mL for NT-proBNP (using $DNN_{Class}^{NT-proBNP}$); and <40 pg/mL, 40–1000 pg/mL, or >1000 pg/mL for cTnI (using $DNN_{Class}^{cTnI}$). Then, in the second stage, quantitative models were used to quantify the corresponding biomarker concentrations within each range, specifically when concentrations in this range exceeded the clinical cut-off level for that biomarker. In contrast, if the predicted classification range entirely fell below the clinical cut-off level, the sample was labeled "negative" for this biomarker without any further quantification. Specifically, samples classified into the <500 pg/mL range by $DNN_{Class}^{CK-MB}$ were labeled "CK-MB negative". Likewise, samples classified into the <125 pg/mL range for NT-proBNP were labeled "NT-proBNP negative". For any other concentration range and biomarker (i.e., ≥500 pg/mL for CK-MB; ≥125 pg/mL for NT-proBNP; <40 pg/mL, 40–1000 pg/mL, and >1000 pg/mL for cTnI), a dedicated quantification model was used to quantify biomarker concentration in this range, resulting in a total of five quantification neural networks ($DNN_{\geq 500}^{CK-MB}$, $DNN_{\geq 125}^{NT-proBNP}$, $DNN_{<40}^{cTnI}$, $DNN_{40-1000}^{cTnI}$, and $DNN_{>1000}^{cTnI}$) that predicted biomarker concentrations in pg/mL. For quality assurance, biomarker concentration predictions from the quantification models were cross-checked with the initial range predicted by the classification model. In a case of a mismatch between the predicted biomarker concentration and the predicted classification range (i.e., predicted concentration fell more than 5% outside the classification range boundaries), the sample was labeled as "*undetermined*" and excluded from the quantification pipeline for a given biomarker. Quantification only occurred when the classification and quantification models agreed with each other, which is used as a computational quality assurance condition.

$DNN_{Class}^{CK-MB}$ was trained on 129 samples from 56 patients and 10 synthetic serum samples (CK-MB spiked in human serum) and blindly tested on a separate dataset of 88 samples from 35 patients and 9 synthetic serum samples. Since $DNN_{\geq 500}^{CK-MB}$ only processes samples with concentrations above 500 pg/mL as determined by $DNN_{Class}^{CK-MB}$, it was trained and tested exclusively on a high-concentration subset of the data. Specifically, the training dataset for $DNN_{\geq 500}^{CK-MB}$ contained 37 samples from 9 patients and 10 synthetic serum samples, while the testing dataset consisted of 28 samples from 6 patients and 9 synthetic serum samples. We supplemented the dataset with spiked serum samples, distributing them evenly across the training and testing sets to ensure effective model learning across all concentration ranges.

As for NT-proBNP, $DNN_{Class}^{NT-proBNP}$ was trained on 109 samples from 57 patients and blindly tested on 70 samples from 35 patients. The subsequent quantification model, $DNN_{\geq 125}^{NT-proBNP}$, was trained on 61 samples from 33 patients and blindly tested on 33 samples from 19 patients that were assigned to the ≥125 pg/mL range by $DNN_{Class}^{NT-proBNP}$.

Furthermore, $DNN_{Class}^{cTnI}$ was trained and tested on the same patients as the $DNN_{Class}^{NT-proBNP}$ model. The training dataset comprised 109 samples from 57 patients. The testing dataset initially contained 70 samples; however, two were removed after failing computational quality assurance between the classification and quantification models, resulting in a final test set of 68 samples from 35 patients. 52 samples from 26 patients were used for the training of $DNN_{<40}^{cTnI}$ model, 35 samples from 19 patients were employed for the training of $DNN_{40-1000}^{cTnI}$ model, and 22 samples from 12 patients were applied for the training of $DNN_{>1000}^{cTnI}$ model. The blind testing datasets contained 30 samples from 15 patients, 24 samples from 12 patients, and 16 samples from 8 patients for $DNN_{<40}^{cTnI}$, $DNN_{40-1000}^{cTnI}$, and $DNN_{>1000}^{cTnI}$ models, respectively.

The biomarker classification models used in the first stage of each test were shallow, fully connected neural networks. The network for CK-MB consisted of three hidden layers (128, 64, 32 units), while the networks for NT-proBNP (64, 32 units) and cTnI (128, 64 units) had two layers each. All hidden layers employed batch normalization (batch size, $N_b$ = 4), a ReLU activation function and dropout. The dropout rates were 0.6 for CK-MB and 0.4 for both NT-proBNP and cTnI. A uniform L2 regularization (α=1e−3) was applied across all networks. The inputs to each neural network were normalized to zero mean and unit variance. Initial learning rates were set to 1e−3 for CK-MB, 1e−2 for NT-proBNP, and 3e−3 for cTnI, and were attenuated by a factor of 0.99 every 10 epochs. The models were optimized using the Adam optimizer with a standard categorical cross-entropy loss function ($L_{CCE}$):

$$L_{CCE} = -\frac{1}{N_b}\sum_{n=1}^{N_b}\sum_{c=1}^{C} y_{n,c} \log y'_{n,c}, \tag{2}$$

where $y_{n,c}$ represents the gold standard binary labels of the $C$ concentration range classes per sample ($C$=2 for CK-MB and NT-proBNP, and $C$=3 for cTnI), and the $y'_{n,c}$ stands for the model's predicted probabilities for these classes, calculated using the softmax activation function as follows:

$$y'_{n,c} = \frac{e^{\hat{y}_{n,c}}}{\sum_{c=1}^{C} e^{\hat{y}_{n,c}}}, \tag{3}$$

where $\hat{y}_{n,c}$ is the model's inference for the class $c$ before the softmax function.

Similarly to the classification networks, the quantification networks also represented shallow, fully connected neural networks, each with two hidden layers. Three cTnI models ($DNN_{<40}^{cTnI}$, $DNN_{40-1000}^{cTnI}$, and $DNN_{>1000}^{cTnI}$) were designed with [128, 64] units, and CK-MB ($DNN_{\geq 500}^{CK-MB}$) and NT-proBNP models ($DNN_{\geq 125}^{NT-proBNP}$) both used [128, 32] units. Each hidden layer utilized a ReLU activation function followed by a dropout layer, with rates set to 0.3 for CK-MB, 0.4 for NT-proBNP, and 0.1 for cTnI. Prior to training, all input features were normalized to have zero mean and unit variance, while the ground truth target concentrations were also transformed: cTnI targets were normalized (zero mean, unit variance), whereas CK-MB and NT-proBNP targets were scaled by factors of 500 and 125, respectively. The models were trained using the Adam optimizer with a batch size ($N_b$) of 20 and an initial learning rate of 1e-3, which was reduced by a factor of 0.5 every 50 epochs. The choice of the loss function was biomarker-dependent: Mean Square Error loss ($L_{MSE}$) for the CK-MB and cTnI models, and Root Mean Square Log Error loss ($L_{RMSLE}$) for the NT-proBNP model:

$$L_{MSE} = \frac{1}{N_b}\sum_{n=1}^{N_b}(y_n - y'_n)^2, \tag{4}$$

$$L_{RMSLE} = \frac{1}{N_b}\sum_{n=1}^{N_b}\sqrt{(\log(y_n + 1) - \log(y'_n + 1))^2}, \tag{5}$$

where $y_n$ represents the ground truth concentration values and $y'_n$ stands for the concentration values predicted by the quantification neural networks. The final model was selected using an early-stopping mechanism. Training was terminated if the loss did not decrease by at least 1e-4 over 200 consecutive epochs, and the model that achieved the lowest loss during training was selected for blind testing.

These models were implemented in Python (v3.8.0) using the PyTorch library (v1.12.1), with training executed on a desktop computer equipped with an Intel Core i9-10920X CPU, 256 GB of RAM, and an NVIDIA GeForce RTX 3090 GPU.

**Feature selection:** To optimize the input features for each network model, we employed an iterative backward feature elimination process[37,39,41]. The procedure started with model training on the complete feature set (i.e., 5 for CK-MB and NT-proBNP models from the colorimetric modality [$\bar{X}^{cTnI}_{Color}$, $\bar{X}^{NT\text{-}proBNP}_{Color}$, $\bar{X}^{CK\text{-}MB}_{Color}$, $\bar{X}^{(+)Ctrl}_{Color}$, $\bar{X}^{(-)Ctrl}_{Color}$] and 6 for cTnI models from both the colorimetric and CL modalities [$\bar{X}^{cTnI}_{Color}$, $\bar{X}^{(+)Ctrl}_{Color}$, $\bar{X}^{(-)Ctrl}_{Color}$, $\bar{X}^{cTnI}_{CL}$, $\bar{X}^{(+)Ctrl}_{CL}$, $\bar{X}^{(-)Ctrl}_{CL}$]). In each subsequent iteration, one feature was systematically removed from the current best-performing subset, and the models were evaluated on training datasets. The subsets achieving the highest accuracy (for classification models) or the lowest MSE/RMSLE loss (for quantification models) were carried forward to the next round. This process was repeated until subsets with a single feature remained. The optimal set of features for each model was determined based on the subset that achieved the highest overall performance. The final optimal feature sets for all models are summarized in **Table 2**, while **Fig. S5** presents a detailed comparison of model performance across test feature sets with different condition numbers during the feature selection process.

**Statistical analysis**: For spiking experiments involving CK-MB, NT-proBNP, and cTnI, results were reported as mean ± SD from at least three independent replicates. Clinical sample measurements were reported as the mean of two replicates ± SD. The number of replicates for each experiment is specified in the corresponding figure captions. For comparing the groups, we used an unpaired two-sample *t*-test, with statistical significance set at $P < 0.05$. The CV was calculated as the SD divided by the mean, expressed as a percentage.

For cTnI, ground truth values obtained from UCLA Health using the hs-cTnI analyzer were directly used. For CK-MB and NT-proBNP, duplicate measurements were performed using a Luminex-based immunoassay and an ELISA, respectively, and the average values were adopted as the ground truth concentrations. Across all clinical samples, the validation assays showed mean CVs of 6.1% for CK-MB and 10.2% for NT-proBNP.

**Table 1.** Comparison of the dual-mode xVFA optical sensor platform with existing commercial and research devices for multiplexed cardiac biomarker testing.

| Ref. | Platform (Category) | Sensing modality | Multiplexing capability per single assay | Specimen (Volume) | LoDs [pg/mL] | Assay ranges [ng/mL] | Cover a clinically relevant range (for cTnI; 0.01–100 ng/mL) | Precision; CV (%) | Assay time | Types of reader device | Reader specs | |
|---|---|---|---|---|---|---|---|---|---|---|---|---|
| | | | | | | | | | | | Size (L× W × D, cm) | Weight (kg) |
| 44 | Abbott Alinity i (Commercial) | Chemiluminescence (CL) | 1-plex per assay (Fragmented) | Serum, plasma (N/C [a]) | cTnI: 0.9<br>NT-proBNP: 7.9<br>Myoglobin (Myo): 12000 | cTnI:0.0027–3.6<br>NT-proBNP: 0.016–35<br>Myo: 23–370 | Insufficient (at high range) | cTnI: 4.0 | 18 min | Benchtop | 134 × 119 × 117 | 624 |
| 45 | Beckman Coulter ACCESS 2 (Commercial) | CL | 1-plex per assay (Fragmented) | Serum, plasma (55 µL per biomarker) | cTnI: 2.0<br>Myo: 762<br>CK-MB: 20<br>NT-proBNP: 10 | cTnI: 0.0021–27.027<br>Myo: 1–4000<br>CK-MB: 0.2–300<br>NT-proBNP: 0.01–35 | Insufficient (at high range) | cTnI: 5.1–6.2 | 17 min | Benchtop | 50 × 99 × 61 | 91 |
| 46 | Polymedco PATHFAST (Commercial) | CL | 1-plex per assay (Fragmented) | Whole blood, plasma (100 µL per biomarker) | cTnI: 2.33<br>Myo: 5000<br>CK-MB: 2000<br>D-Dimer: 5000<br>NT-proBNP: 15<br>hs-CRP: 50 | cTnI: 0.0023–50<br>Myo: 5–1000<br>CK-MB: 2–500<br>D-Dimer: 5–5000<br>NT-proBNP: 0.015–30<br>hs-CRP: 50–30000 | Insufficient (at high ranges) | cTnI: <7.1<br>Myo: <5<br>CK-MB: <9<br>D-Dimer: <7.1<br>NT-proBNP: <6<br>hsCRP: <9 | <17 min | Benchtop | 75 × 57 × 51 | 33 |
| 47 | Abbott i-STAT (Commercial) | Electrochemical | 1-plex per assay (Fragmented) | Whole blood, plasma (~22 µL per biomarker) | hs-cTnI: 1.61<br>cTnI: 20<br>CK-MB: 0.6<br>BNP: 15 | hs-cTnI: 0.0029–1<br>cTnI: 0–50<br>CK-MB: 0–150<br>BNP: 0.015–5 | Insufficient (at high range) | hs-cTnI: <18% | ~15 min | Portable | 7.7 × 23.5 × 7.3 | 0.65 |
| 48 | Siemens Atellica VTLi (Commercial) | Optical magnetic | No | Whole blood, capillary blood, plasma (30–100 µL) | cTnI: 1.24 | cTnI: 0.00124–1.25 | Insufficient (at high range) | cTnI: 6.7 | 8 min | Portable | 25 × 8.5 × 5.2 | 0.78 |
| 49 | Electrode array (Research) | Electrochemical | 2-plex | Serum (<50 µL) | cTnI: 1<br>cTnT: 1 | cTnI: 0.0001–100<br>cTnT: 0.0001–100 | Sufficient | <10% | ~2 h | Benchtop | 19 × 9 × 27 | 3 |
| 50 | Electrode array (Research) | Electrochemilumines ce (ECL) | 3-plex | Serum (2.5 µL for CCD /100 µL for PMT) | Myo: 31<br>cTnI: 0.79<br>cTnT: 300 | Myo:<br>0.050–1.0 (PMT); 0.5–10 (CCD)<br>cTnI:<br>0.0010–0.010 (PMT); 0.5–10 (CCD)<br>cTnT:<br>0.50–4.0 (PMT); 5.0–100 (CCD) | Insufficient (at high range) | N/C | >2 h | Benchtop | N/C | N/C |
| 51 | Electrode array (Research) | Electrical & Dielectrophoresis | 1-plex per assay (Fragmented) | Buffer (20 µL) | cTnI: 100<br>cTnT: 100 | cTnI: 0.05–0.8<br>cTnT: 0.05–0.8 | Insufficient (at low and high ranges) | N/C | 2 min | Portable | N/C | N/C |
| 52 | Electrode array (Research) | Electrochemical | 4-plex | Plasma, whole blood (15 µL) | cTnI: 24<br>NT-proBNP: 3<br>BNP: N/C | cTnI: 0.01–10<br>NT-proBNP: 0.01–10<br>BNP: 0.01–10 | Insufficient (at low and high ranges) | N/C | >40 min | Benchtop | N/C | N/C |

| | | | | | | | | | | | |
|---|---|---|---|---|---|---|---|---|---|---|---|
| 53 | Electrode array (Research) | ECL | 3-plex | Diluted serum (60 µL; 10- or 100-fold)) | cTnI: 0.024<br>h-FABP: 0.053<br>copeptin: 0.014 | cTnI: 0.0001–1<br>f-FABP: 0.0001–1<br>copeptin: 0.0001–1 | Insufficient (at high range) | N/C | >40 min | Benchtop | N/C | N/C |
| 32 | Plasmonic gold microarray (Research) | Fluorescence | 2-plex | Diluted serum (100 µL) | cTnI: 10<br>CK-MB: 250 | cTnI: 0.01–1.2<br>CK-MB: 0.25–64 | Insufficient (at high range) | <15% | 30–150 min | Benchtop | 36.9 × 27.8 × 45.7 | 15.5 |
| 54 | Optofluidic sensor (Research) | Evanescent wave interactions-based detection | 3-plex | Serum | Myo: 500<br>cTnI: 2.6<br>CK-MB: 3.7 | Myo: 2–2000<br>cTnI: 0.02–20<br>CK-MB: 0.2–200 | Insufficient (at high range) | N/C | >20 min | Benchtop | N/C | N/C |
| 55 | SERS on plasmonic metasurface (Research) | Surface Enhanced Raman Scattering (SERS) | 3-plex | Serum (Volume: N/C) | cTnI: 7<br>CK-MB: 50<br>Myo: 3800 | cTnI: 0.01–9<br>CK-MB: 0.3–33.3<br>Myo: 3.3–1333 | Insufficient (at high range) | N/C | 25 min | Benchtop | N/C | N/C |
| 56 | Field-effect transistor (FET) with microfluidics (Research) | Electrochemical | 4-plex | Serum (4 µL) | cTnI: 0.394<br>NT-proBNP: 0.832<br>CRP: 140,000<br>Fibrinogen: $2.02\times10^{8}$ | cTnI: 0.001–10<br>NT-proBNP: 0.05–10<br>CRP: 100–50,000<br>Fibrinogen: 100,000–5,000,000 | Insufficient (at high range) | N/C | 5 min | Portable | 2 compartments: 25 × 12 × 15 & 30 × 25 × 22 | N/C |
| 57 | Digital microfluidics (Research) | Fluorescence | 3-plex | Serum (~6 µL) | Myo: 3<br>cTnI: 10.5<br>CK-MB: 300 | Myo: 25–1000<br>cTnI: 0.15–10<br>CK-MB: 5–100 | Insufficient (at high range) | N/C | 30 min | Benchtop | N/C | N/C |
| 58 | Aptamer-based electrode microfluidics (Research) | Electrochemical | 4-plex | Whole blood (20 µL) | cTnI: 0.54<br>NT-proBNP: 1.53<br>Fibrinogen: $5.94\times10^{8}$<br>CRP: $3.9\times10^{5}$ | cTnI: 0.0001–10<br>NT-proBNP: 0.0001–10<br>Fibrinogen: $5\times10^{5}$–$1\times10^{7}$<br>CRP: 500–9000 | Insufficient (at high range) | N/C | 15 min | Benchtop | N/C | N/C |
| 59 | Electrochemical paper-based microfluidics (Research) | Electrochemical | 3-plex | Serum (4 µL per biomarker) | CRP: 380<br>cTnI: 0.16<br>PCT: 0.27 | CRP: 1–$10^{5}$<br>cTnI: 0.001–250<br>PCT: 0.0005–250 | Sufficient | N/C | ~45 min | Benchtop | N/C | N/C |
| 60 | Spin exchange relaxation-free-based microfluidics (Research) | Magnetometer | 4-plex | Buffer (200 µL per biomarker) | cTnI: 10<br>BNP: 10<br>H-FABP: 100<br>CRP: 100 | cTnI: 0.01–10<br>BNP: 0.01–10<br>H-FABP: 0.1–100<br>CRP: 0.1–1000 | Insufficient (at high range) | ~5% | ~30 min | Benchtop | N/C | N/C |
| 61 | Paper-based lateral flow assay (LFA) (Research) | SERS | 3-plex | Serum (100 µL) | Myo: 3.2<br>cTnI: 0.44<br>CK-MB: 0.55 | Myo: 0.01–500<br>cTnI: 0.01–50<br>CK-MB: 0.02–90 | Sufficient | 8.5% | >20 min | Benchtop | 61.0 × 160.0 × 61.0 | 90 |
| 62 | Paper microfluidics (Research) | SERS | 3-plex | Serum (10 µL) | GPBB: 8<br>CK-MB: 10<br>cTnT: 1 | GPBB: 0.1–100<br>CK-MB: 0.1–100<br>cTnT: 0.01–200 | Not relevant | <10% | ~27 min | Benchtop | N/C | N/C |
| 63 | Paper-based LFA<br>(Research) | Magnetic | 3-plex | Serum<br>(80 µL) | Myo: 50<br>cTnI: 8.9<br>CK-MB: 63 | Myo: 0.17–1000<br>cTnI: 0.03–250<br>CK-MB: 0.21-–250 | Insufficient<br>(at low range) | N/C | ~13 min | Benchtop | N/C | N/C |
| 64 | Paper-based LFA<br>(Research) | Fluorescence | 3-plex | Serum<br>(80 µL) | Myo:2000<br>cTnI: 50<br>CK-MB:200 | Myo: 2.0–1000<br>cTnI: 0.05–25<br>CK-MB: 0.2–100 | Insufficient<br>(at low and high ranges) | Myo< 10%<br>cTnI: <15%<br>CK-MB< 10% | >17 min | Portable | N/C | N/C |
| 65 | Paper-based LFA<br>(Research) | Fluorescence | 3-plex | Diluted plasma or serum<br>(10 µL) | Myo: 1000<br>cTnI: 10<br>CK-MB:1000 | Myo: 1–1000<br>cTnI: 0.1–50<br>CK-MB:0.5–500 | Insufficient<br>(at low range) | <8% | 12 min | Benchtop &<br>Portable | 22.0 × 12.0 × 10.0 | N/C |
| 66 | Paper-based LFA<br>(Research) | Fluorescence | 3-plex | Serum<br>(60 µL) | cTnI: 36<br>Myo: 540<br>CK-MB: 250 | cTnI: 0.12–125<br>Myo: 5–640<br>CK-MB: 1.5–192 | Insufficient<br>(at low range) | <8% | 10 min | Portable | N/C | N/C |
| 67 | Paper microfluidics<br>(Research) | Fluorescence | 3-plex | Serum<br>(15 µL) | Myo: 2380<br>cTnI: 1000<br>h-FABP: 1360 | Myo: 5–60<br>cTnI: 1–50<br>h-FABP: 2.5–60 | Insufficient<br>(at low range) | N/C | 5 min | Portable | N/C | N/C |
| 68 | Paper microfluidics<br>(Research) | CL | 3-plex | Serum<br>(100 µL) | cTnI: 0.50<br>h-FABP: 0.32<br>copeptin: 0.40 | cTnI: 0.001–1<br>h-FABP: 0.001–1<br>copeptin: 0.001–1 | Insufficient<br>(at high range) | N/C | >40 min | Benchtop | N/C | N/C |
| 69 | Paper microfluidics<br>(Research) | CL | 3-plex | Serum<br>(2.5 µL per biomarker) | cTnI: 0.3<br>h-FABP: 0.06<br>copeptin: 0.4 | cTnI: 0.0005–1000<br>h-FABP: 0.0001–1000<br>copeptin: 0.001–1,000,000 | Sufficient | N/C | >30 min | Benchtop | N/C | N/C |
| 70 | Paper microfluidics<br>(Research) | Colorimetric | 3-plex | Serum<br>(10 µL) | cTnT: 50<br>CK-MB: 500<br>GPBB: 500 | cTnT: 0.05–200<br>CK-MB: 0.5–100<br>GPBB: 0.5–100 | Not relevant | 10–20% | 10 min | Portable | N/C | N/C |
| 71 | Paper-based electrophoretic assay<br>(Research) | Colorimetric | 3-plex | Serum<br>(500 µL per biomarker) | cTnI: 12.1<br>HDL: $4.36\times10^{8}$<br>LDL: $3.83\times10^{8}$ | cTnI: 0.0001–0.1<br>HDL: $1\times10^{5}$–$5\times10^{5}$<br>LDL: $1\times10^{5}$–$1\times10^{6}$ | Insufficient<br>(at high range) | N/C | <6 min | Portable | N/C | N/C |
| 39 | Paper-based vertical flow assay (VFA)<br>(Research) | Fluorescence | 3-plex | Serum<br>(50 µL) | Myo: 520<br>CK-MB: 300<br>h-FABP: 490 | Myo: 0.52–75<br>CK-MB: 0.30–23.4<br>h-FABP: 0.49–45.7 | Not relevant | Myo: 12.4%<br>CK-MB: 12.6%<br>h-FABP: 12.5 % | <15 min | Portable | 16 × 8 × 4.5 | 0.32 |
| **This Work** | Paper-based VFA<br>(Research) | Colorimetric and CL | 3-plex | Serum<br>(50 µL) | cTnI: 0.12<br>NT-proBNP: 40<br>CK-MB: 409 | cTnI: 1–100<br>NT-proBNP: 0.1–100<br>CK-MB: 1–100 | Sufficient | <5% | 23 min | Portable | 15.5 × 10.5× 14.0 | 0.7 |

[a] "N/C" denotes "not commented" on the corresponding reference.

**Table 2**. Neural network models and their corresponding optimal feature sets.

| Models | All available features | Optimal feature sets |
|---|---|---|
| $DNN^{CK-MB}_{Class}$ | $[\bar{X}^{cTnI}_{Color}, \bar{X}^{NT\text{-}proBNP}_{Color}, \bar{X}^{CK\text{-}MB}_{Color}, \bar{X}^{(+)Ctrl}_{Color}, \bar{X}^{(-)Ctrl}_{Color}]$ | $[\bar{X}^{cTnI}_{Color}, \bar{X}^{NT\text{-}proBNP}_{Color}, \bar{X}^{CK\text{-}MB}_{Color}, \bar{X}^{(+)Ctrl}_{Color}, \bar{X}^{(-)Ctrl}_{Color}]$ |
| $DNN^{NT-proBNP}_{Class}$ | $[\bar{X}^{cTnI}_{Color}, \bar{X}^{NT\text{-}proBNP}_{Color}, \bar{X}^{CK\text{-}MB}_{Color}, \bar{X}^{(+)Ctrl}_{Color}, \bar{X}^{(-)Ctrl}_{Color}]$ | $[\bar{X}^{NT\text{-}proBNP}_{Color}, \bar{X}^{CK\text{-}MB}_{Color}, \bar{X}^{(+)Ctrl}_{Color}]$ |
| $DNN^{cTnI}_{Class}$ | $[\bar{X}^{cTnI}_{Color}, \bar{X}^{(+)Ctrl}_{Color}, \bar{X}^{(-)Ctrl}_{Color}, \bar{X}^{cTnI}_{CL}, \bar{X}^{(+)Ctrl}_{CL}, \bar{X}^{(-)Ctrl}_{CL}]$ | $[\bar{X}^{cTnI}_{Color}, \bar{X}^{(+)Ctrl}_{Color}, \bar{X}^{(-)Ctrl}_{Color}, \bar{X}^{cTnI}_{CL}, \bar{X}^{(+)Ctrl}_{CL}]$ |
| $DNN^{CK-MB}_{\geq 500}$ | $[\bar{X}^{cTnI}_{Color}, \bar{X}^{NT\text{-}proBNP}_{Color}, \bar{X}^{CK\text{-}MB}_{Color}, \bar{X}^{(+)Ctrl}_{Color}, \bar{X}^{(-)Ctrl}_{Color}]$ | $[\bar{X}^{CK\text{-}MB}_{Color}]$ |
| $DNN^{NT-proBNP}_{\geq 125}$ | $[\bar{X}^{cTnI}_{Color}, \bar{X}^{NT\text{-}proBNP}_{Color}, \bar{X}^{CK\text{-}MB}_{Color}, \bar{X}^{(+)Ctrl}_{Color}, \bar{X}^{(-)Ctrl}_{Color}]$ | $[\bar{X}^{cTnI}_{Color}, \bar{X}^{NT\text{-}proBNP}_{Color}, \bar{X}^{CK\text{-}MB}_{Color}, \bar{X}^{(-)Ctrl}_{Color}]$ |
| $DNN^{cTnI}_{<40}$ | $[\bar{X}^{cTnI}_{Color}, \bar{X}^{(+)Ctrl}_{Color}, \bar{X}^{(-)Ctrl}_{Color}, \bar{X}^{cTnI}_{CL}, \bar{X}^{(+)Ctrl}_{CL}, \bar{X}^{(-)Ctrl}_{CL}]$ | $[\bar{X}^{cTnI}_{CL}, \bar{X}^{(-)Ctrl}_{CL}]$ |
| $DNN^{cTnI}_{40-1000}$ | $[\bar{X}^{cTnI}_{Color}, \bar{X}^{(+)Ctrl}_{Color}, \bar{X}^{(-)Ctrl}_{Color}, \bar{X}^{cTnI}_{CL}, \bar{X}^{(+)Ctrl}_{CL}, \bar{X}^{(-)Ctrl}_{CL}]$ | $[\bar{X}^{cTnI}_{CL}]$ |
| $DNN^{cTnI}_{>1000}$ | $[\bar{X}^{cTnI}_{Color}, \bar{X}^{(+)Ctrl}_{Color}, \bar{X}^{(-)Ctrl}_{Color}, \bar{X}^{cTnI}_{CL}, \bar{X}^{(+)Ctrl}_{CL}, \bar{X}^{(-)Ctrl}_{CL}]$ | $[\bar{X}^{cTnI}_{Color}, \bar{X}^{(+)Ctrl}_{Color}, \bar{X}^{(-)Ctrl}_{Color}, \bar{X}^{cTnI}_{CL}, \bar{X}^{(-)Ctrl}_{CL}]$ |

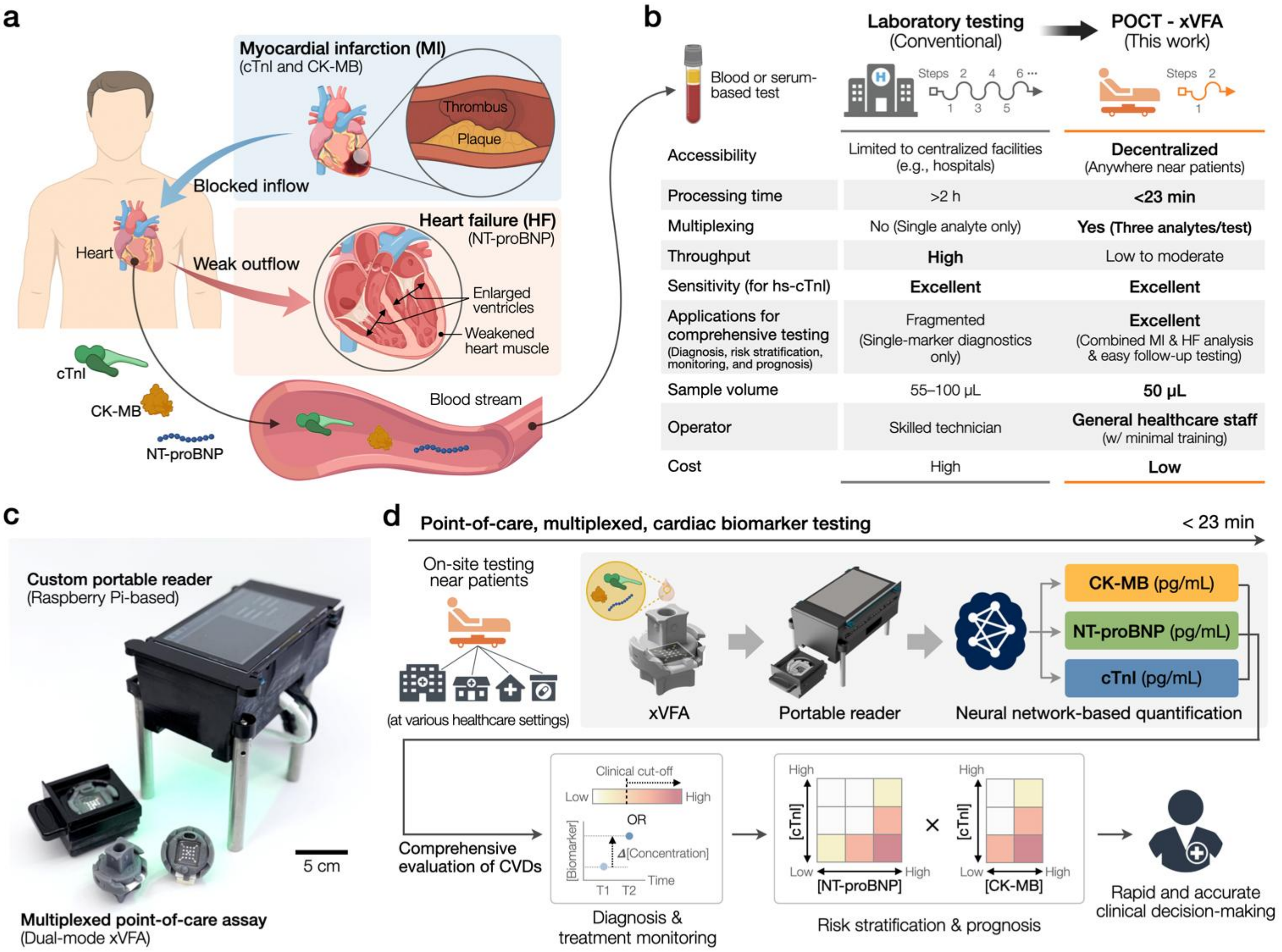

**Fig. 1.** Overview of the dual-mode xVFA optical sensor platform for multiplexed cardiac biomarker detection in point-of-care settings. (a) Pathophysiological relationship between MI and HF, and their associated biomarkers: cTnI and CK-MB for MI, and NT-proBNP for HF. (b) Comparison between conventional central laboratory testing and the proposed point-of-care dual-mode xVFA optical sensor. (c) Image of the dual-mode xVFA system, including assay/imaging cartridges and a custom portable imaging-based optical reader (Raspberry Pi-based). (d) Workflow of deep learning-powered dual-mode xVFA. The integrated outputs can support a comprehensive evaluation of cardiovascular diseases (CVDs), including diagnosis, treatment monitoring, risk stratification, and prognosis, facilitating rapid and accurate clinical decisions.

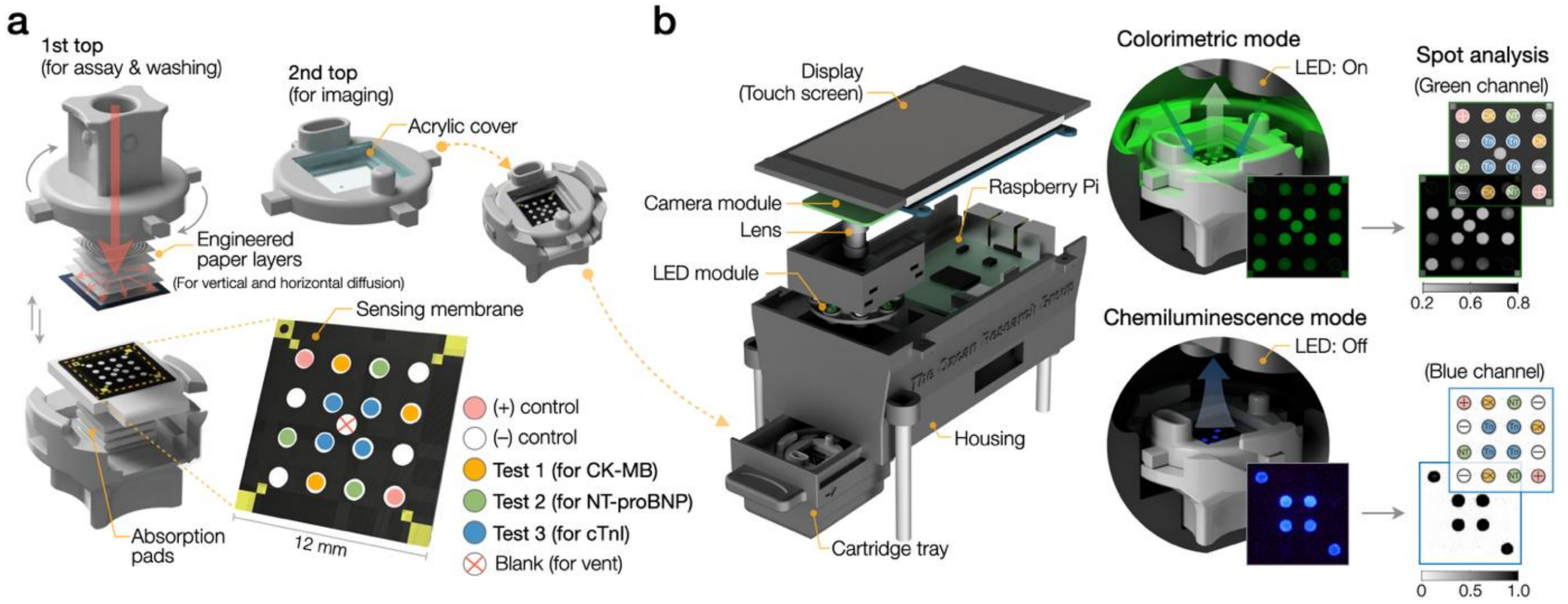

**Fig. 2.** Overview of the dual-mode xVFA sensor and portable optical reader. (a) Design of the dual-mode xVFA cartridge. The first top case is for reagent delivery and washing, while the second top case is for imaging. The bottom case features a multiplexed sensing membrane with 16 immunoreaction spots, including test spots for CK-MB, NT-proBNP, and cTnI, as well as internal positive and negative controls. (b) Structure of the Raspberry Pi-based portable optical reader and dual-mode imaging setup. Colorimetric signals are captured under green LED illumination (LED on), whereas CL signals are acquired with the LED off. The captured images are processed to identify spots and extract the optical signal intensity.

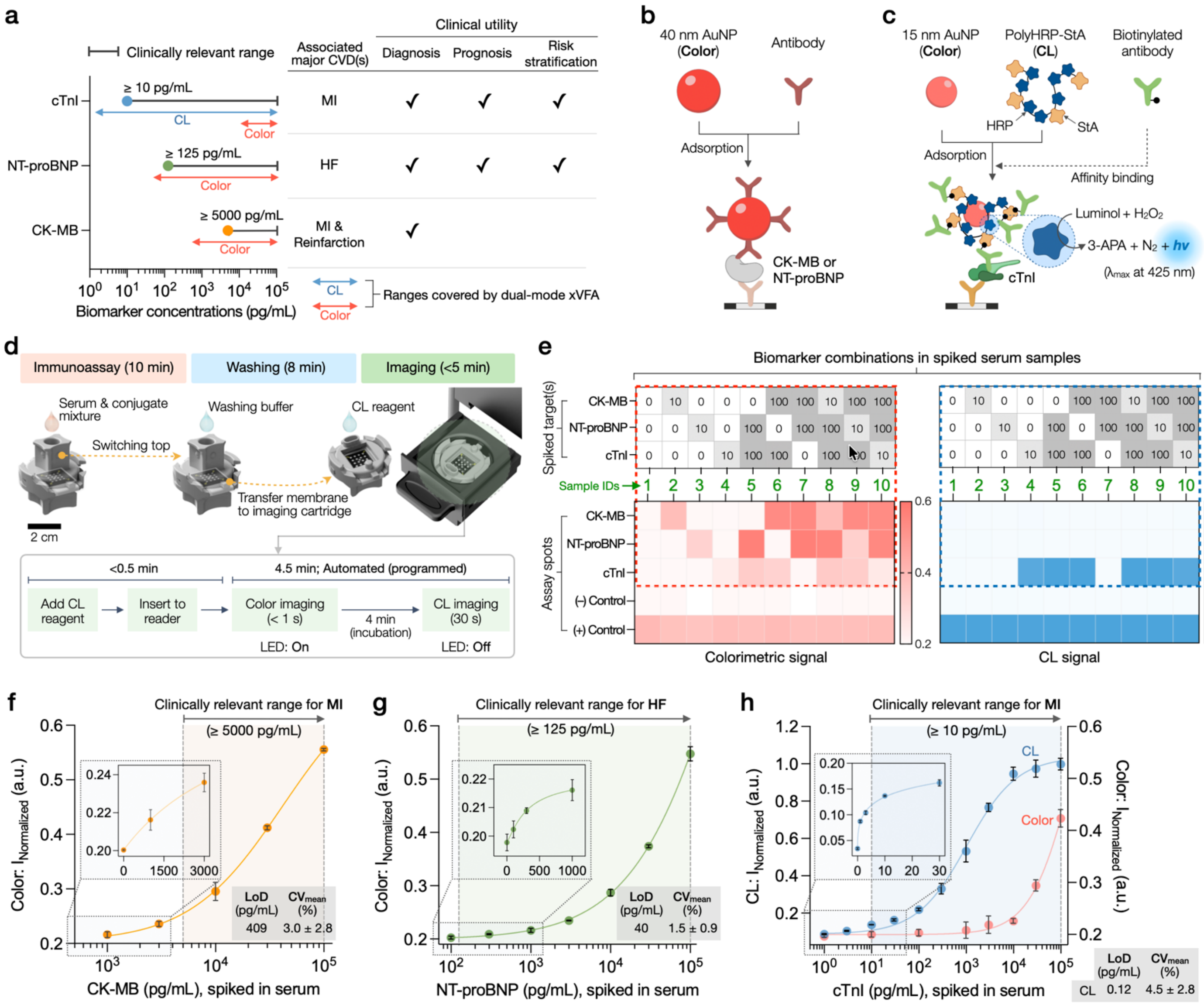

**Fig. 3.** Multiplexed cardiac biomarker detection enabled by dual-mode optical sensing in xVFA. (a) Biomarker-to-modality mapping for three cardiac biomarkers (cTnI, NT-proBNP, and CK-MB) targeted in this work, highlighting their clinically relevant concentration ranges, associated CVDs, and clinical utility. Blue and red arrows represent the measurable ranges of the dual-mode xVFA platform using CL and colorimetric modalities, respectively. (b) Sensing conjugate and mechanism of colorimetric detection for CK-MB and NT-proBNP. (c) Sensing conjugate and mechanism of colorimetric and CL detection for cTnI. PolyHRP-StA refers to polymerized horseradish peroxidase conjugated with streptavidin. (d) Assay procedure of the dual-mode xVFA. (e) Cross-reactivity evaluation results of the multiplexed dual-mode assay. Top heatmaps represent the spiked concentrations (ng/mL) and combinations of CK-MB, NT-proBNP, and cTnI across ten serum samples. Bottom heatmaps show the corresponding colorimetric and CL signal intensities measured from the assay spots for each sample. The green numbered columns (1–10) indicate spiked sample IDs, each representing a unique combination of biomarker concentrations. Dashed rectangles visually link the input concentration matrices (top) to their respective signal outputs (bottom), illustrating how the multiplexed assay responds to different biomarker combinations. Calibration curves and detection performance for (f) CK-MB ($y=0.0002x^{0.6901}$, $R^2=0.997$), (g) NT-proBNP ($y=0.0001x^{0.6752}$, $R^2=0.998$), and (h) cTnI ($y=0.0742x^{0.2744}$, $R^2=0.996$ for CL and $y=0.0179x^{0.2739}$, $R^2=0.996$ for colorimetric), were established using serum samples spiked with known concentrations of each biomarker. Data points in (f)–(h) represent the mean of triplicates ± SD. Vertical dashed lines in (f)–(h) indicate the clinically relevant cut-offs of each biomarker.

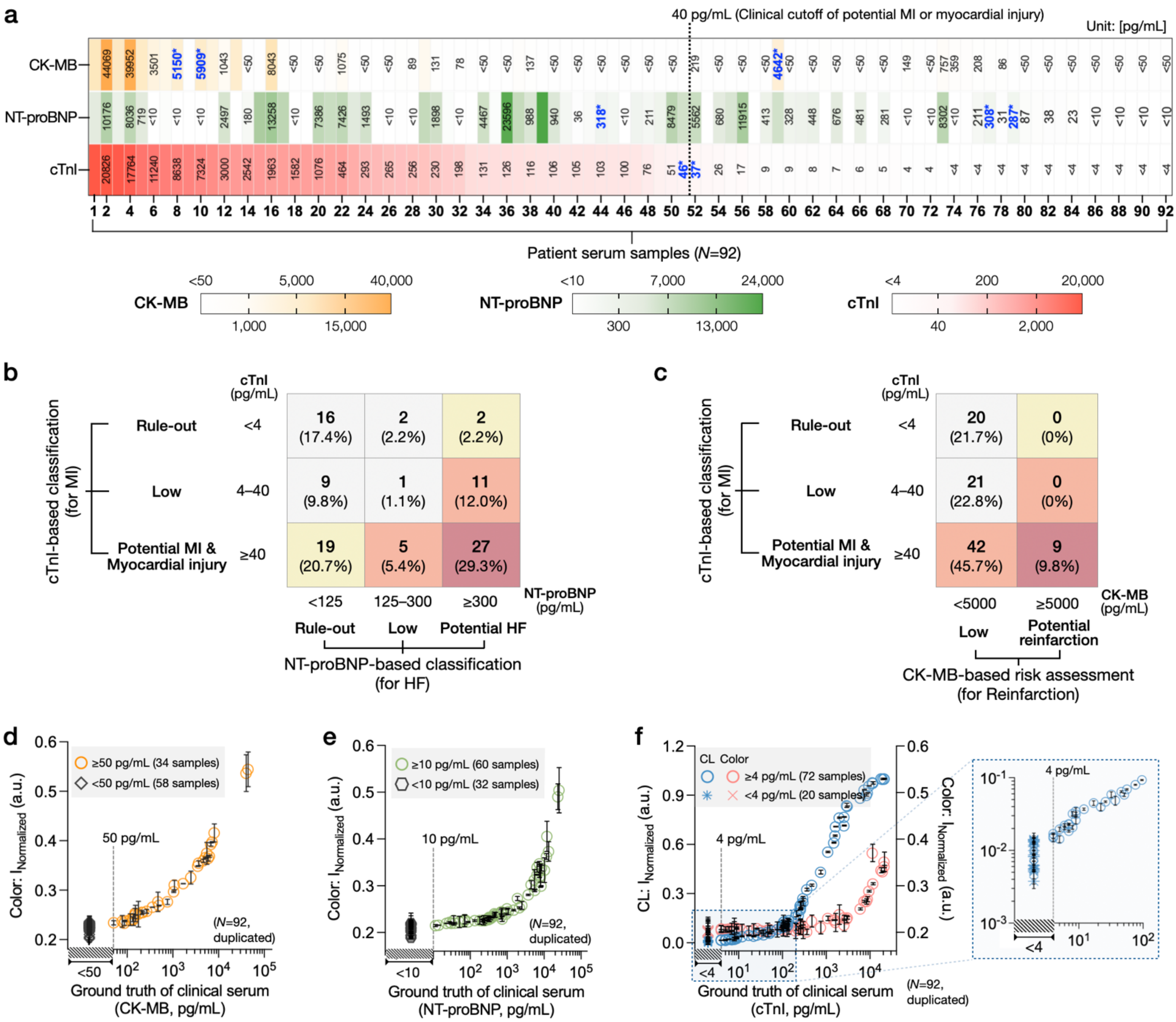

**Fig. 4.** Distribution and classification of CK-MB, NT-proBNP, and cTnI levels in clinical serum samples. (a) Ground truth distribution summary in 92 patients' serum samples. Samples are arranged in descending order based on ground truth cTnI concentrations. Vertical dashed line between samples 51 and 52 indicates clinically relevant cut-off, 40 pg/mL for cTnI (potential myocardial injury or infarction). Blue bold values with an asterisk (*) mark the concentrations near the clinical cut-off levels of each biomarker. (b) Cross-classification of clinical serum samples based on cTnI and NT-proBNP levels, illustrating the distribution of patients across MI- and HF-related risk categories. Colors reflect increasing severity, from rule-out (light) to potential MI and HF (dark). (c) Cross-classification of serum samples based on cTnI and CK-MB levels, highlighting the potential identification of reinfarction. (d)–(f) Validation of dual-mode xVFA measurements using 92 clinical human serum samples with ground truth concentrations: (d) CK-MB, (e) NT-proBNP, and (f) cTnI. Vertical dashed lines in (d)–(f) indicate the cut-offs of the assays used to measure the ground truth concentrations of each biomarker; 50 pg/mL for CK-MB (from Luminex assay), 10 pg/mL for NT-proBNP (from ELISA), and 4 pg/mL for cTnI (from standard analyzer; Access 2 from Beckman Coulter). Samples below each cut-off concentration do not have a quantitative ground truth, and the x-axis does not display quantitative values in that range. All samples below the cut-off levels are stacked on top of each other within the diagonal hatching interval. Data points in (d)–(f) represent the mean of duplicates ± SD. Insets show magnified plots at low concentration ranges.

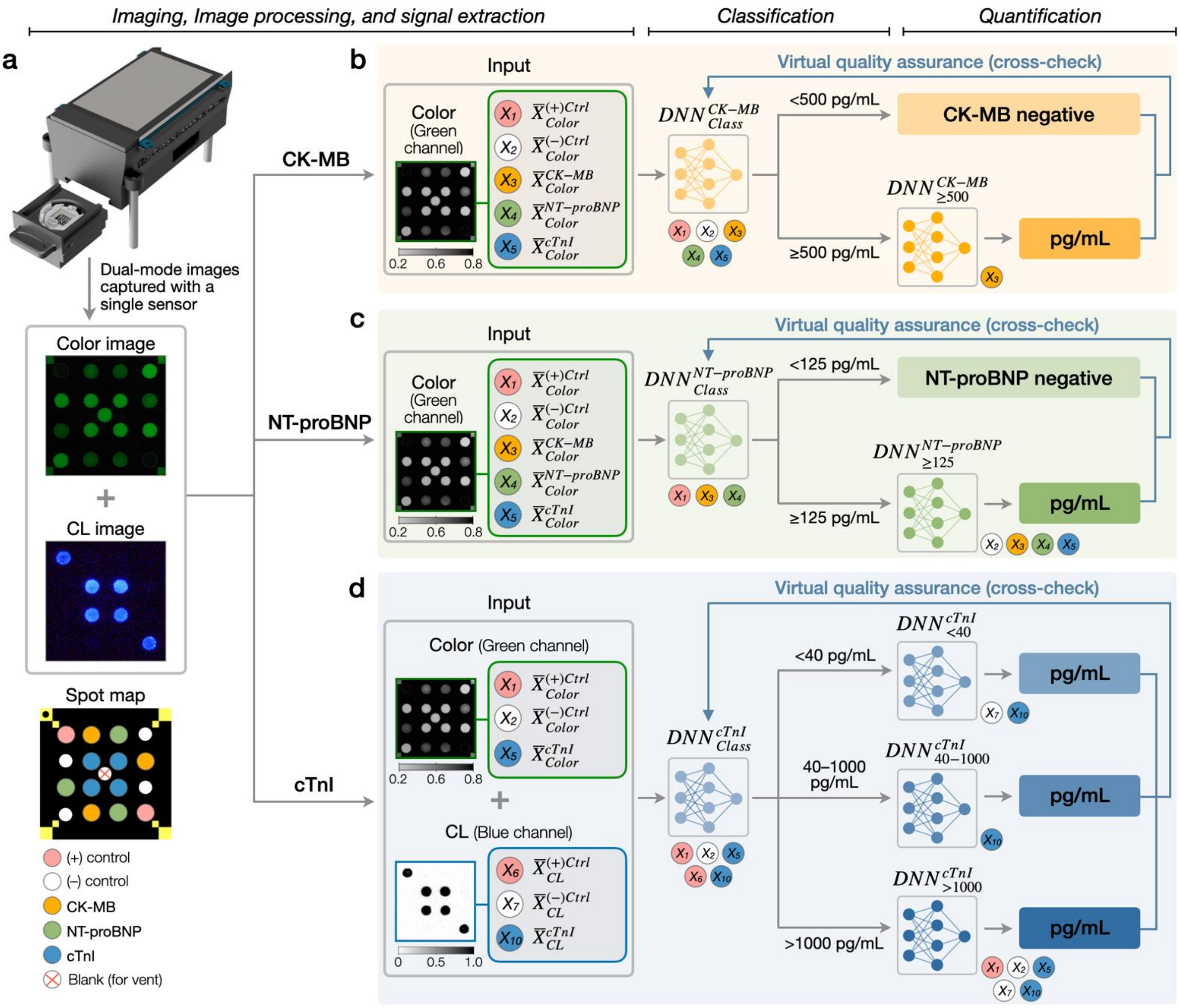

**Fig. 5.** Neural network-based quantification pipelines for the 3 target biomarkers (CK-MB, NT-proBNP, and cTnI). (a) Representative colorimetric and CL images of the dual-mode xVFA from a single optical sensor captured by the portable imaging-based reader (top) and spot map layout identifying each test and control position (bottom). (b) Deep learning-based CK-MB quantification pipeline, consisting of 1 classification ($\mathrm{DNN}^{\mathrm{CK-MB}}_{\mathrm{Class}}$) and 1 quantification ($\mathrm{DNN}^{\mathrm{CK-MB}}_{\geq 500}$) neural networks. First, $\mathrm{DNN}^{\mathrm{CK-MB}}_{\mathrm{Class}}$ uses colorimetric signals to classify samples between <500 pg/mL and ≥500 pg/mL CK-MB concentration ranges. Only samples classified into the ≥500 pg/mL range are further processed by $\mathrm{DNN}^{\mathrm{CK-MB}}_{\geq 500}$ for CK-MB concentration inference, while samples in the <500 pg/mL range are reported as CK-MB negative. (c) Deep learning-based NT-proBNP quantification pipeline, consisting of 1 classification ($\mathrm{DNN}^{\mathrm{NT-proBNP}}_{\mathrm{Class}}$) and 1 quantification ($\mathrm{DNN}^{\mathrm{NT-proBNP}}_{\geq 125}$) networks. $\mathrm{DNN}^{\mathrm{NT-proBNP}}_{\mathrm{Class}}$ uses colorimetric signals to classify samples between <125 pg/mL and ≥125 pg/mL NT-proBNP concentration ranges. Only samples in the ≥125 pg/mL range are processed by $\mathrm{DNN}^{\mathrm{NT-proBNP}}_{\geq 125}$ for NT-proBNP concentration inference, while samples in the <125 pg/mL range are reported as NT-proBNP negative. (d) Deep learning-based cTnI quantification pipeline, consisting of 1 classification ($\mathrm{DNN}^{\mathrm{cTnI}}_{\mathrm{Class}}$) and 3 quantification ($\mathrm{DNN}^{\mathrm{cTnI}}_{<40}$, $\mathrm{DNN}^{\mathrm{cTnI}}_{40\text{-}1000}$, and $\mathrm{DNN}^{\mathrm{cTnI}}_{>1000}$) neural networks. $\mathrm{DNN}^{\mathrm{cTnI}}_{\mathrm{Class}}$ uses both colorimetric and CL signals to classify each sample between three cTnI concentration ranges: <40 pg/mL, 40–1000 pg/mL, or >1000 pg/mL. Then, three separate quantification network models ($\mathrm{DNN}^{\mathrm{cTnI}}_{<40}$, $\mathrm{DNN}^{\mathrm{cTnI}}_{40\text{-}1000}$, and $\mathrm{DNN}^{\mathrm{cTnI}}_{>1000}$) quantify cTnI concentration in samples within each range. For each biomarker, all quantification results are cross-checked with classification results. If any discrepancy occurs between the classification and quantification inference, the sample is labeled as "undetermined" for a given biomarker.

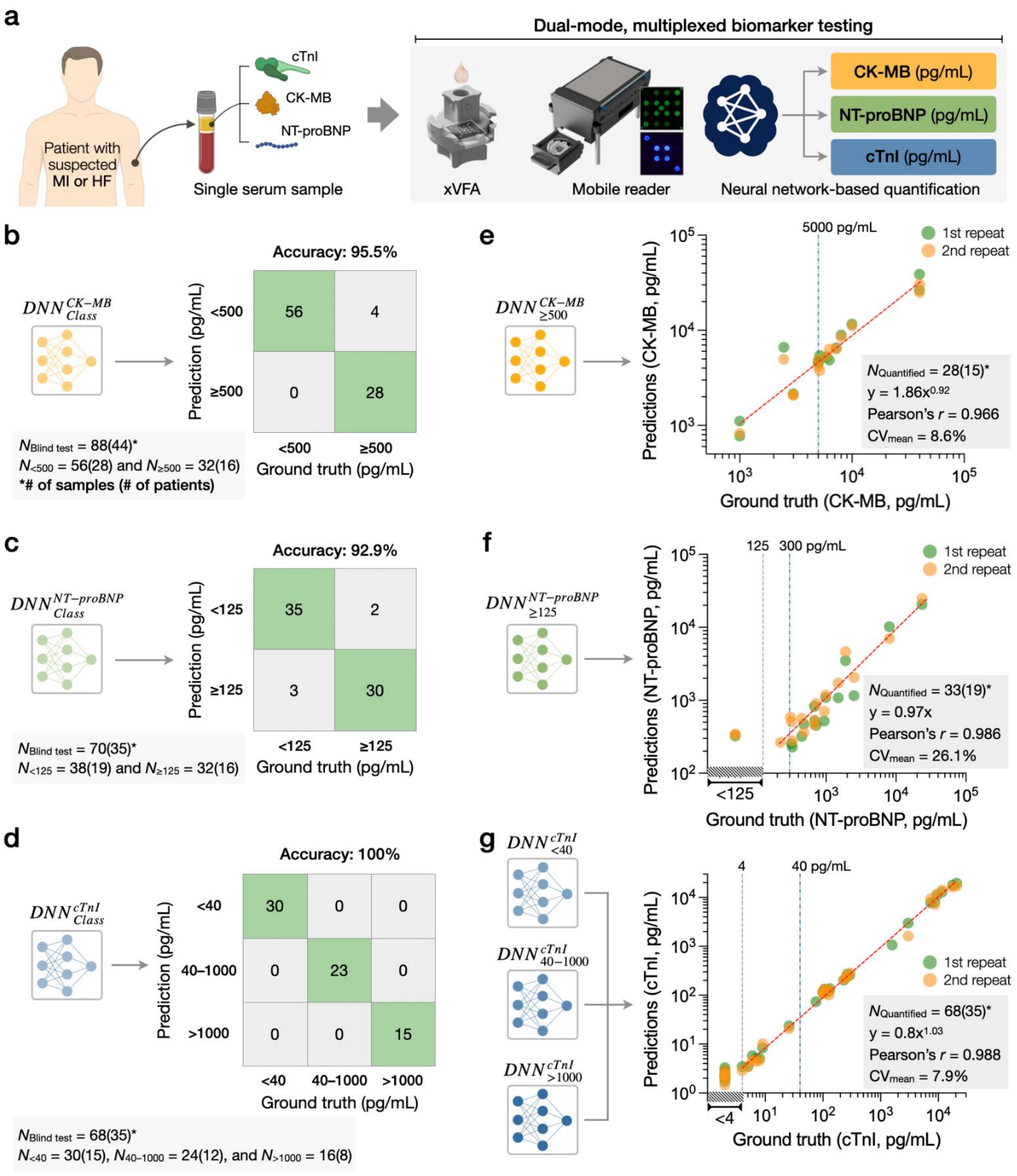

**Fig. 6.** Neural network-based analysis of the 3 cardiac biomarkers measured by the dual-mode xVFA optical sensor. (a) Schematic of the overall testing pipeline: a single serum sample is tested using the dual-mode optical xVFA, and both colorimetric and CL signals are captured using a portable reader. The resulting signals are then analyzed via neural network models for biomarker classification and quantification. (b) Classification results of $\mathrm{DNN}^{\mathrm{CK-MB}}_{\mathrm{Class}}$ for 88 samples from 35 patients and 9 synthetic serums (CK-MB spiked in serum) used in the blind testing set. (c) Classification results of $\mathrm{DNN}^{\mathrm{NT-proBNP}}_{\mathrm{Class}}$ for 70 samples from 35 patients used in the blind testing set. (d) Classification results of $\mathrm{DNN}^{\mathrm{cTnI}}_{\mathrm{Class}}$ for 68 samples from 35 patients used in the blind testing set. (e)

Quantification results of $\text{DNN}_{\geq 500}^{\text{CK-MB}}$ for 28 samples from 6 patients and 9 synthetic serums classified into the ≥500 pg/mL CK-MB concentration range by $\text{DNN}_{\text{Class}}^{\text{CK-MB}}$. Samples classified into <500 pg/mL range by $\text{DNN}_{\text{Class}}^{\text{CK-MB}}$ were labeled as CK-MB negative and were not processed by the quantification model. Vertical dashed line at 5000 pg/mL indicates the commonly used clinical cut-off for CK-MB, above which MI or reinfarction is suspected. (f) Quantification results of $\text{DNN}_{\geq 125}^{\text{NT-proBNP}}$ for 33 samples from 19 patients classified into the ≥125 pg/mL NT-proBNP concentration range by $\text{DNN}_{\text{Class}}^{\text{NT-proBNP}}$. Samples classified into <125 pg/mL range by $\text{DNN}_{\text{Class}}^{\text{NT-proBNP}}$ were labeled as NT-proBNP negative and were not processed by the quantification model. Vertical dashed lines at 125 pg/mL and 300 pg/mL indicate clinical reference thresholds: 125 pg/mL is commonly used as a rule-out cut-off for chronic HF, while 300 pg/mL is frequently employed in diagnosing acute HF. (g) Combined quantification results from the three cTnI quantification models ($\text{DNN}_{<40}^{\text{cTnI}}$, $\text{DNN}_{40\text{-}1000}^{\text{cTnI}}$, and $\text{DNN}_{>1000}^{\text{cTnI}}$) for 68 samples from 35 patients used in the blind testing set. 2 samples (i.e., ground truth of 76 pg/mL and 1582 pg/mL) were excluded from quantification due to a discrepancy found during cross-check between classification and quantification results. Samples with the cTnI concentration in <4 pg/mL range do not have a quantitative ground truth (due to the reporting limit of the FDA-approved analyzer), and therefore x-axis does not have quantitative values in that range. All samples from the blind testing set that fall in the <4 pg/mL range are stacked above each other within the <4 pg/mL interval. Vertical dashed lines at 4 pg/mL and 40 pg/mL indicate relevant clinical thresholds: 4 pg/mL represents the reporting limit of a standard laboratory analyzer used for ground truth measurements, while 40 pg/mL is a proposed decision threshold for detecting myocardial injury with high-sensitivity cTnI assays.